\documentclass[sigconf]{acmart}

\usepackage{subfigure}
\usepackage{multirow}
\usepackage{appendix}
\usepackage{listings}
\usepackage{tcolorbox}
\usepackage{color}

\ccsdesc[500]{Software and its engineering~Software reverse engineering}
\ccsdesc[500]{Software and its engineering~Search-based software engineering}

\copyrightyear{2024}
\acmYear{2024}
\setcopyright{acmlicensed}\acmConference[ICSE '24]{2024 IEEE/ACM 46th International Conference on Software Engineering}{April 14--20, 2024}{Lisbon, Portugal}
\acmBooktitle{2024 IEEE/ACM 46th International Conference on Software Engineering (ICSE '24), April 14--20, 2024, Lisbon, Portugal}
\acmDOI{10.1145/3597503.3639080}
\acmISBN{979-8-4007-0217-4/24/04}

\definecolor{revision}{rgb}{0,0,0}
\definecolor{revision2}{rgb}{0,0,0}

\newcommand{\inlined}{Leaf-Inlining}
\newcommand{\inlining}{Root-Inlining}
\newcommand{\recursive}{Internal-Inlining}

\newcommand{\inlinedshort}{Leaf}
\newcommand{\inliningshort}{Root}
\newcommand{\recursiveshort}{Internal}

\begin{document}

\title{Cross-Inlining Binary Function Similarity Detection}


\author{Ang Jia}
\email{jiaang@stu.xjtu.edu.cn}
\affiliation{
	\institution{Xi'an Jiaotong University}
	\country{China}
}

\author{Ming Fan}
\email{mingfan@mail.xjtu.edu.cn}
\affiliation{
	\institution{Xi'an Jiaotong University}
	\country{China}
}
\authornote{Corresponding author}

\author{Xi Xu}
\email{xx19960325@stu.xjtu.edu.cn}
\affiliation{
	\institution{Xi'an Jiaotong University}
	\country{China}
}

\author{Wuxia Jin}
\email{jinwuxia@mail.xjtu.edu.cn}
\affiliation{
	\institution{Xi'an Jiaotong University}
	\country{China}
}
\author{Haijun Wang}
\email{haijunwang@xjtu.edu.cn}
\affiliation{
	\institution{Xi'an Jiaotong University}
	\country{China}
}

\author{Ting Liu}
\email{tingliu@mail.xjtu.edu.cn}
\affiliation{
	\institution{Xi'an Jiaotong University}
	\country{China}
}

\begin{abstract}
	Binary function similarity detection plays an important role in a wide range of security applications. Existing works usually assume that the query function and target function share equal semantics and compare their full semantics to obtain the similarity. However, we find that the function mapping is more complex, especially when function inlining happens.
	
	In this paper, we will systematically investigate cross-inlining binary function similarity detection. We first construct a cross-inlining dataset by compiling 51 projects using 9 compilers, with 4 optimizations, to 6 architectures, with 2 inlining flags, which results in two datasets both with 216 combinations. Then we construct the cross-inlining function mappings by linking the common source functions in these two datasets. Through analysis of this dataset, we find that three cross-inlining patterns widely exist while existing work suffers when detecting cross-inlining binary function similarity. Next, we propose a pattern-based model named CI-Detector for cross-inlining matching. CI-Detector uses the attributed CFG to represent the semantics of binary functions and GNN to embed binary functions into vectors.  CI-Detector respectively trains a model for these three cross-inlining patterns. Finally, the testing pairs are input to these three models and all the produced similarities are aggregated to produce the final similarity. We conduct several experiments to evaluate CI-Detector. Results show that CI-Detector can detect cross-inlining pairs with a precision of 81\% and a recall of 97\%, which exceeds all state-of-the-art works.
\end{abstract}


\keywords{Cross-Inlining, Binary Similarity Detection, Inlining Pattern}

\maketitle

\section{Introduction}

Most software today is not developed entirely from scratch. Instead, developers rely on a range of open-source components to create their applications~\cite{kula2018developers}. According to a recent report~\cite{CodeResue}, 96\% of the software contains open-source code. Although using open-source components helps to finish projects quicker and reduce costs, improper reuse introduces security and legal risks~\cite{gkortzis2021software}. Of the 1,703 codebases scanned in 2022, 87\% include security and operational risk assessments~\cite{CodeResue}, where 54\% of software has license conflicts and 84\% of software contains at least one vulnerability. To make it worse, downstream software often relies on close-sourced third-party libraries~\cite{TPL}. Security and legal risks hidden in the binaries generated by the upstream supplier may be unintentionally transferred to the downstream developers or end-users.

To help resolve these code-reuse-related issues, \textcolor{revision2}{many} binary code similarity analysis works are proposed and have been applied in various applications, including code search~\cite{asm2vec, Kam1n0, safe, CodeCMR, Genius, bingo-E}, OSS reuse detection~\cite{BAT, RESource, JISIS14, BinPro, OSSPolice, Saner2019, B2SFinder, ban2021b2smatcher, ISRD}, vulnerability detection~\cite{Tracy, Gemini, vulseeker, Firmup} and patch presence test~\cite{FIBER, Binxray, PG-VulNet, PDiff, Osprey}. They usually regard the vulnerable functions or reused functions as the query functions and the functions in the commercial software as the target functions and produce the detection results by calculating the similarity between query functions and target functions.

To obtain function-to-function similarities, most existing binary similarity analysis works usually assume that the target function shares the same semantics with the query function and try to find exact matches between them. However, we discover that the query function and the target function do not always share equal semantics, especially when function inlining happens. 

\begin{figure}[t]
	\centering
	\subfigure[do\_free\_upto]{
		\begin{minipage}[t]{0.33\linewidth}
			\centering
			\includegraphics[width=1\textwidth]{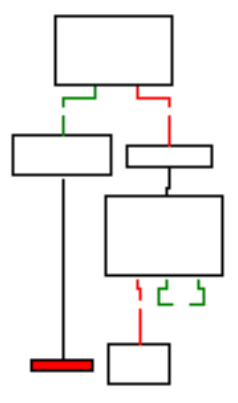}
			\label{fig:motivating_example_a}
			\vspace{-10pt}
		\end{minipage}%
	}%
	\centering
	\subfigure[CMS\_final]{
		\begin{minipage}[t]{0.33\linewidth}
			\centering
			\includegraphics[width=1\textwidth]{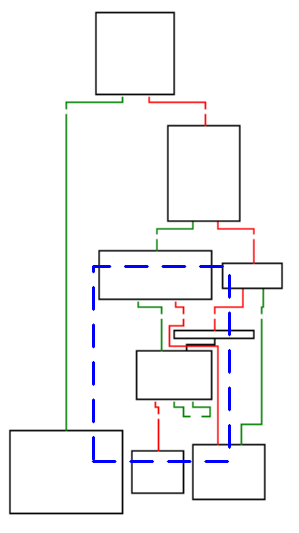}
			\label{fig:motivating_example_b}
			\vspace{-10pt}
		\end{minipage}%
	}%
	\subfigure[CMS\_decrypt]{
		\begin{minipage}[t]{0.33\linewidth}
			\centering
			\includegraphics[width=1\textwidth]{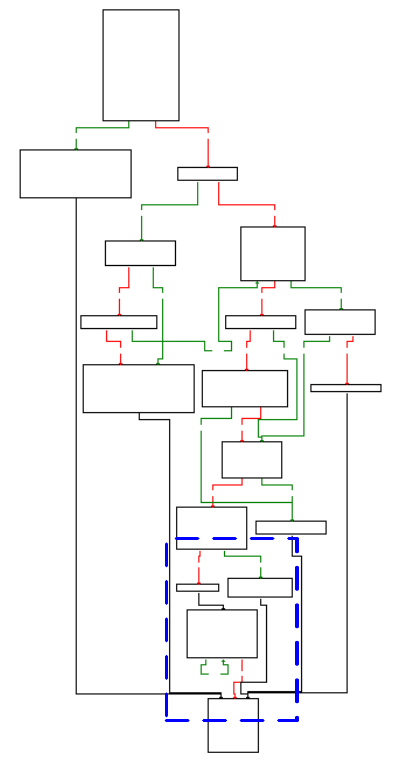}
			\label{fig:motivating_example_c}
			\vspace{-10pt}
		\end{minipage}%
	}%
	\vspace{-15pt}
	\caption{Cross-inlining matching example}
	\label{fig:motivating_example}
	\vspace{-12pt}
\end{figure}

Figure~\ref{fig:motivating_example} shows a vulnerable binary function \textit{do\_free\_upto} and two binary functions \textit{CMS\_final} and \textit{CMS\_decrypt} in OpenSSL 1.0.1m~\cite{openssl} compiled using gcc-8.2.0 with O3. Figure~\ref{fig:motivating_example_a} shows the CFG (control flow graph) of the vulnerable function \textit{do\_free\_upto}. 
\textit{do\_free\_upto} is associated with the CVE-2015-1792~\cite{CVE-2015-1792} which lacks an examination of null value that allows remote attackers to cause a denial of service. Figure~\ref{fig:motivating_example_b} and Figure~\ref{fig:motivating_example_c} respectively presents CFG of function \textit{CMS\_final} and \textit{CMS\_decrypt}. From the debug information, we notice that function \textit{CMS\_final} and \textit{CMS\_decrypt} both inline the vulnerable function \textit{do\_free\_upto}, copy the function body of \textit{do\_free\_upto} (we use the blue dotted rectangle to represent it), and thus inherit the vulnerability.

However, when we try to use the existing works~\cite{Tracy, Gemini, vulseeker, Firmup} to match the vulnerable function \textit{do\_free\_upto} with the function \textit{CMS\_final} and \textit{CMS\_decrypt}, \textcolor{revision}{the returned similarities are all less than 50\%}. We notice that though functions \textit{CMS\_final} and \textit{CMS\_decrypt} are vulnerable, they also contain function contents from their own and other inlined functions. \textcolor{revision}{For example, binary function \textit{CMS\_final} is compiled from source function \textit{CMS\_final} and \textit{do\_free\_upto}, while binary function \textit{CMS\_decrypt} is compiled by source function \textit{CMS\_decrypt}, \textit{check\_content} and \textit{do\_free\_upto}.} The vulnerable function is only part of the function \textit{CMS\_final} and \textit{CMS\_decrypt}. Thus, it is difficult to identify the vulnerable function through exact matching.


Function inlining widely exists in binaries~\cite{jia20221}. Though many works~\cite{asm2vec, Kam1n0, safe, CodeCMR, Genius, bingo-E, ISRD, Tracy, Gemini, vulseeker, Firmup, FIBER, Binxray, PG-VulNet, PDiff, Osprey} are proposed to resolve the cross-optimization, cross-compiler, and cross-architecture problems in binary code similarity, few works are targeted at cross-inlining.
There are still three challenges for resolving the cross-inlining binary function similarity detection.

\textbf{C1: Obscure binary semantics.} Different from source code which has rich semantics and is easy to understand, the semantics in binaries is obscure and hard to separate. Moreover, when function inlining happens, the semantics of several functions are mixed, which makes it harder to separate the inlined functions.

\textbf{C2: Different inlining contexts.} The query functions (vulnerable function or reused function) can be called and inlined by different functions. For example, the vulnerable function \textit{do\_free\_upto} has been inlined into five different functions in one single binary.
The different contexts around the inlined functions lead to different compositions of final functions.

\textbf{C3: Various inlining patterns.} Apart from the cases in which a vulnerable function is inlined into other functions, we also noticed that a vulnerable function can also inline other functions. For example, the vulnerable function \textit{ssl23\_get\_client\_hello} in OpenSSL 1.0.1j, associated with CVE-2014-3569~\cite{CVE-2014-3569}, has inlined another normal function \textit{ssl2\_get\_server\_method} when compiling using gcc-8.2.0 with O3. The various inlining patterns make cross-inlining detection a more complicated problem. 


To tackle the challenges that function inlining brings, in this paper, we propose the first study to systematically investigate the solution of cross-inlining binary function similarity detection. 

\textcolor{revision}{To tackle \textbf{C1}, we combine the opcodes and CFG to form ACFG (Attributed CFG) to represent the function semantics. As shown in Figure~\ref{fig:motivating_example}, though the vulnerable function is inlined into other functions, the vulnerable part has a similar structure and similar code logic. }

To tackle \textbf{C2}, we use GNN~\cite{GMN_matching} to learn the similarity of cross-inlining function pairs. 
GNN can learn from local structures of ACFG and thus identify the similar part of cross-inlining function pairs.

To tackle \textbf{C3}, we first conduct an empirical study and summarize three kinds of inlining patterns. Then, we train separate models for each inlining pattern to calculate cross-inlining similarities. Finally, we aggregate the similarities from these models to obtain the final detection results. 

We evaluate our method on a cross-inlining dataset and results show that our method can obtain an AUC of 90.62\% and exceed all state-of-the-art methods.

\textcolor{revision}{We summarize our contribution as follows}:

\begin{itemize}
	\item We create a cross-inlining dataset and we propose a labeling method to identify cross-inlining function pairs. 
	
	\item We conduct an empirical study on the cross-inlining dataset and we summarize three inlining patterns that help better learn the cross-inlining similarities.
	
	\item Based on the three patterns, we propose a pattern-based ensemble method named CI-Detector. CI-Detector can detect cross-inlining pairs with a precision of 81\% and a recall of 97\%, which exceeds all state-of-the-art works.
	
\end{itemize}


To facilitate further research, we have made the source code and dataset publicly available~\cite{github_repo}.

\section{Background}
\label{sec:background}

\subsection{Binary Function Similarity Detection}

There are two major ways to detect binary code similarity, dynamic and static approaches.
Dynamic approaches~\cite{DBLP:conf/icisc/MingPG12, DBLP:conf/acsac/LindorferFMCZ12, DBLP:conf/uss/JangWB13, DBLP:conf/uss/EgeleWCB14, DBLP:conf/sec/MingXW15, DBLP:conf/wcre/HuZLG16, DBLP:journals/tjs/KimLKI19, DBLP:conf/uss/MingXJW17, DBLP:conf/kbse/WangW17, DBLP:conf/kbse/KargenS17}, which obtain the program semantic by executing the program with test input, usually have a high overhead and suffer from low coverage.
Static approaches~\cite{asm2vec, Kam1n0, safe, CodeCMR, Genius, bingo-E, Tracy, Gemini, vulseeker, Firmup}, which directly extract features from binary executables, are more scalable to large amounts of binaries. However, under different compilation settings, the same source code can be compiled into binaries with different syntax, different layouts, and different instruction sets, which introduces cross-optimization, cross-compiler, and cross-architecture binary similarity detection approaches.

Cross-optimization binary similarity detection works~\cite{Genius, bingo, Gemini, alphadiff, vulseeker, RLZ2019, innereye, asm2vec, safe, Tracy, Esh, Gitz, Firmup} aim to detect similarities of binary functions compiled by same source functions but with different optimizations. For example, Gitz~\cite{Gitz} decomposes binary procedures to comparable fragments and uses the compiler to re-optimize them to obtain canonical strands for comparison. Though different optimizations lead to syntactically different binary codes, re-optimizing them with high optimizations converts them into syntactically similar binaries.

Cross-compiler detection works~\cite{Genius, Gemini, bingo, alphadiff, asm2vec, safe} aim to detect similarities of binary functions compiled by different compilers. For example, Asm2Vec~\cite{asm2vec} learns the lexical semantic relationships and constructs the function embedding by aggregating instruction embeddings. Different compilers use different conventions to arrange the binary code while existing works learn the semantics from adjacent instructions and aggregate them to detect similar functions.

Cross-architecture detection works~\cite{Genius, bingo, Gitz, Gemini, alphadiff, vulseeker, RLZ2019, innereye, safe} aim to detect similarities of binary functions compiled to different architectures. For example, Gemini~\cite{Gemini} constructs ACFG (Attributed Control Flow Graph) and leverages graph embedding to generate binary function embeddings. Though different instruction sets are used in different architectures, the control flow structure can still preserve similar execution logic in binary code. 

Though existing works have paid enormous effort to cross-optimization, cross-compiler, and cross-architecture problems, few works have systematically explored cross-inlining binary function similarity detection. Some works including Bingo~\cite{bingo} and Asm2Vec~\cite{asm2vec} have proposed manually designed rules to resolve some issues under function inlining, but they cannot cover the various inlining patterns and cannot provide a complete view of the cross-inlining binary function similarity detection.

\subsection{Function Inlining}

Function inlining, or inline expansion, is a manual or compiler optimization that replaces a function call site with the body of the called function~\cite{wiki_inlining}. Though the direct effect of function inlining is to eliminate call overhead, its primary benefit is to allow further optimizations. As the contents of multiple functions are aggregated together, more intra-procedural optimizations can be applied without requiring inter-procedural optimizations.

Though inlining brings improvement to the executable performance, it also increases code size. Thus, it is a trade-off to determine when to inline for a balance between these benefits and costs. A range of different heuristics~\cite{DBLP:journals/tse/DavidsonH92, DBLP:conf/lfp/DeanC94, hubicka2004gcc, DBLP:conf/lcpc/ZhaoA03, DBLP:conf/cc/CooperHW08, andersson2009evaluation, DBLP:conf/pldi/HwuC89, durand2018partial} have been explored for inlining. Usually, an inlining algorithm has a certain code budget (an allowed increase in program size) and aims to inline the most valuable call sites without exceeding that budget. 

Once the compiler has decided to inline a particular function, performing the inlining operation itself is usually simple. Depending on whether the compiler inlines functions across code in different languages, the compiler can do inlining on either a high-level intermediate representation (like abstract syntax trees) or a low-level intermediate representation. In either case, the compiler simply computes the arguments, stores them in variables corresponding to the function's arguments, and then inserts the body of the function at the call site~\cite{wiki_inlining}.

\subsection{\textcolor{revision}{Cross-Inlining Binary Similarity Detection}}
\label{sec:related_work}

Existing works have taken preliminary research of binary code similarity analysis under function inlining. The first binary function similarity detection work that takes function inlining into consideration is Bingo~\cite{bingo}. Bingo summarizes several patterns that a caller function should inline its callee functions, and \textcolor{revision2}{conducts} inlining to simulate the functions generated by function inlining. Asm2Vec~\cite{asm2vec} also uses Bingo's strategies to tackle function inlining. \textcolor{revision2}{However, inlining the callee function into caller functions requires that the query function and the target function should have the same context in FCG} \textcolor{revision2}{(Function Call Graph)}. It also raises a higher cost to compare functions with bigger function contents.

Jia~\cite{Tosem} has conducted the first systematical empirical study to investigate the effect of function inlining on binary function similarity analysis. They evaluated four works on the dataset with inlining, and results show that most works suffer a 30\%-40\% performance loss when detecting the functions with inlining. Besides, most works ignored the inlined functions, causing functions that have inlined vulnerable functions \textcolor{revision2}{to be undetected}. Our work also noticed these two issues, and we propose a method that can tackle these two challenges.

Recently, Jia~\cite{jia2022comparing} has proposed a method named O2NMatcher to investigate the binary2source matching method under function inlining. In binary2source matching, binary functions can be generated by several source functions. O2NMatcher tries to find the source function sets as the matching targets of binary functions with inlining. That requires the context of the source functions. In this paper, we will propose a more scalable method without requiring the context of query functions and target functions. Besides, we will introduce two additional cross-inlining patterns in this work.

\section{Empirical study}
\label{sec:empirical_study}

In this section, we will conduct an empirical study to investigate cross-inlining binary function similarity detection tasks. 

\subsection{Problem Definition}
\label{sec:defination}
We first define the objective of cross-inlining binary function similarity detection as follows:

Given a query function \textit{q} without inlining and a target function \textit{t} with inlining, in binary form, the cross-inlining binary function similarity detection aims to detect whether function \textit{q} is inlined into the function \textit{t}. 


In this paper, we take the first step in cross-inlining detection --- we study the cross-inlining problems without introducing cross-optimization, cross-compiler, and cross-architecture problems. That is the query function \textit{q} and the target function \textit{t} are compiled by the same compilers using the same optimizations to the same architectures. However, 
\textcolor{revision2}{the query function \textit{q} and the target function \textit{t} can still be compiled}
under multiple optimizations, compilers, and architecture.

\subsection{Cross-inlining Dataset Construction}
\label{sec:dataset}

We construct a cross-inlining dataset by using Binkit~\cite{Binkit} and propose a cross-inlining labeling method leveraging the function inlining identification method in Jia~\cite{Tosem}.

Binkit is a binary code similarity analysis benchmark. We use it to construct two datasets: \textit{Dataset-Inlining} and \textit{Dataset-NoInlining}. Both datasets are constructed by compiling 51 gnu projects~\cite{gnulib} using 9 compilers (GCC v{4.9.4, 5.5.0, 6.4.0, 7.3.0, 8.2.0} and Clang v{4.0, 5.0, 6.0, 7.0}), 4 optimizations (O0, O1, O2, O3), to 6 architectures (ARM32, ARM64, MIPS32, MIPS64, X86-32. X86-64), resulting in total 216 combinations. \textit{Dataset-NoInlining} is compiled with an additional flag \textit{``fno-inline''} to turn off the function inlining. Finally, \textit{Dataset-Inlining} is composed of 50,760 binaries and 12,644,259 binary functions, where 2,195,920 (17.4\%) are binary functions with inlining. \textit{Dataset-NoInlining} is composed of 50,760 binaries and 15,233,501 binary functions.

We further construct the binary2source function mappings for these two datasets respectively. We follow the workflow in Jia~\cite{Tosem}: we first extract the address-to-line mappings from the \textit{.debug\_line} section, then we extract the address-to-binary-function and line-to-source-function relation to get the functions that the address or line belongs to, and finally we construct the binary2source function mappings by aligning the binary functions with their mapped source functions.

Using the binary2source function mappings, we can identify binary functions with inlining by the number of source functions they map. If a binary function maps to more than one source function, it will be considered a function with inlining. In \textit{Dataset-NoInlining}, the binary functions are mostly mapped to one source function. Note that these are binary functions with inlining in \textit{Dataset-NoInlining}, as \textit{``fno-inline''} cannot prevent user-forced inlining such as ``$always\_inline$''. We exclude these binary functions with inlining from \textit{Dataset-NoInlining}.

\begin{figure}[h]
	\centering
	\vspace{-5pt}
	\includegraphics[width=0.44\textwidth]{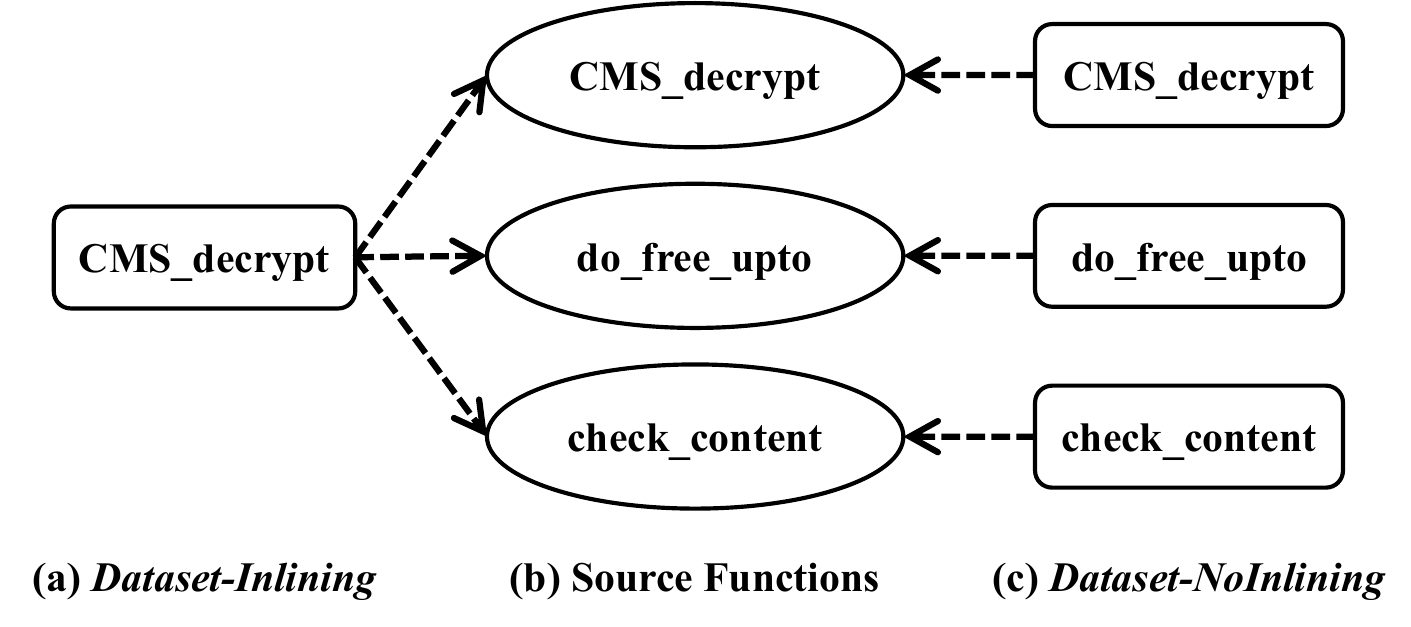}
	\vspace{-8pt}
	\caption{Example of constructing cross-inlining mappings}
	\label{fig:b2b_example}
	\vspace{-8pt}
\end{figure}

Then we use the source function as the bridge to construct cross-inlining function mappings. Figure~\ref{fig:b2b_example} presents an example. We use rounded rectangles to represent binary functions and ellipses to represent source functions. The dotted arrows represent the binary2source function mappings. After constructing binary2source function mappings, we notice that binary function \textit{CMS\_decrypt} in \textit{Dataset-Inlining} maps to source functions \textit{CMS\_decrypt}, \textit{do\_free\_upto}, and \textit{check\_content}. The binary function \textit{do\_free\_upto} in \textit{Dataset-NoInlining} maps to source function \textit{do\_free\_upto}. Here, we regard the source function \textit{do\_free\_upto} as a \textit{bridge function}. Leveraging the bridge function, we can obtain a cross-inlining function pair: \textit{CMS\_decrypt} in \textit{Dataset-Inlining} and \textit{do\_free\_upto} in \textit{Dataset-NoInlining}.

By traversing all binary functions in \textit{Dataset-Inlining} and \textit{Dataset-NoInlining}, we construct 5,621,140 cross-inlining function pairs.

\subsection{Cross-Inlining Patterns Analysis}
\label{sec:patterns}

After constructing the cross-inlining function mappings, we further investigate the cross-inlining patterns and their existence in the cross-inlining dataset. Firstly, we summarize three cross-inlining patterns according to the location of the bridge function in the FCG.

\begin{figure}[t]
	\centering
	\subfigure[\inlined]{
		\begin{minipage}[t]{0.33\linewidth}
			\centering
			\includegraphics[width=1\textwidth]{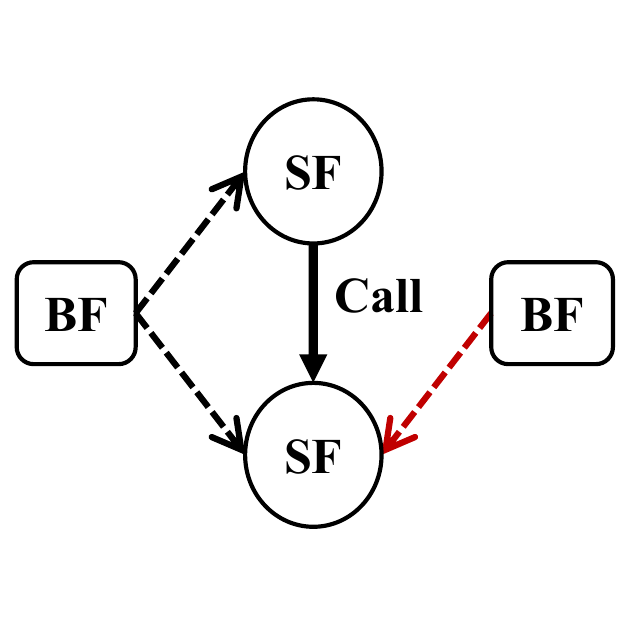}
			\label{fig:matching_pattern_a}
			\vspace{-10pt}
		\end{minipage}%
	}%
	\centering
	\subfigure[\inlining]{
		\begin{minipage}[t]{0.33\linewidth}
			\centering
			\includegraphics[width=1\textwidth]{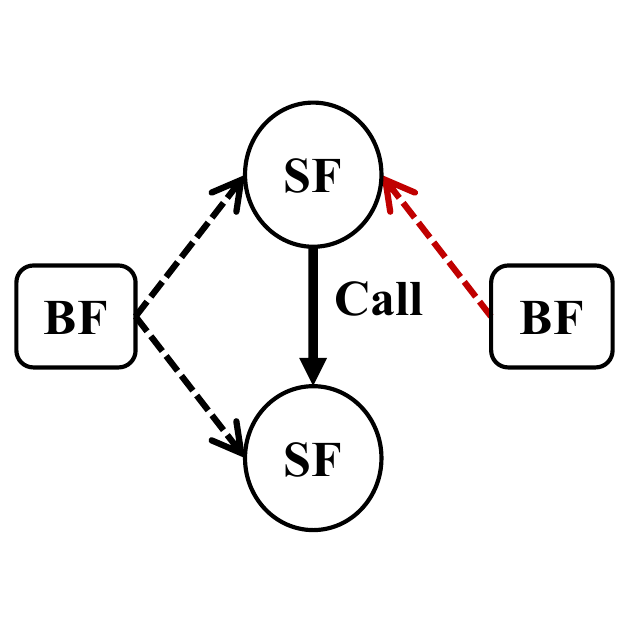}
			\label{fig:matching_pattern_b}
			\vspace{-10pt}
		\end{minipage}%
	}%
	\subfigure[\recursive]{
		\begin{minipage}[t]{0.33\linewidth}
			\centering
			\includegraphics[width=1\textwidth]{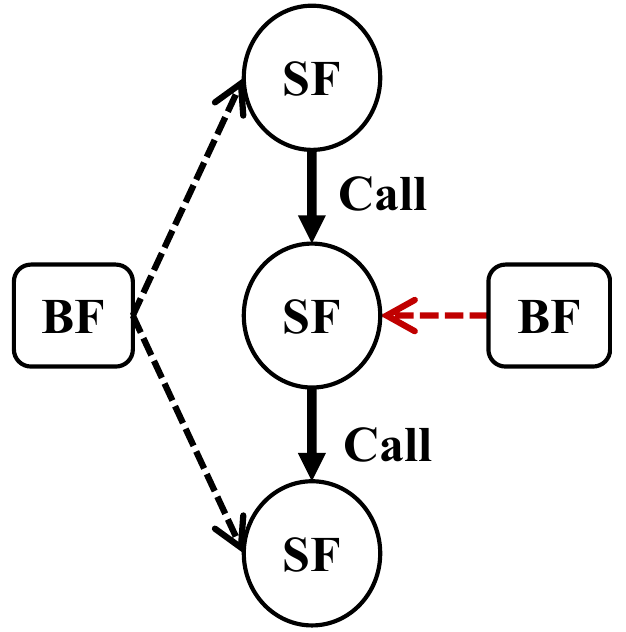}
			\label{fig:matching_pattern_c}
			\vspace{-10pt}
		\end{minipage}%
	}%
	\vspace{-15pt}
	\caption{Cross-inlining matching patterns}
	\label{fig:matching_pattern}
	\vspace{-15pt}
\end{figure}

Figure~\ref{fig:matching_pattern} shows three cross-inlining patterns. We use rounded rectangles to represent binary functions (BF) and circles to represent source functions (SF). We use dotted arrows to represent binary2source function mappings and solid arrows to represent function calls between source functions. We summarize three cross-inlining patterns including \textit{\inlined}, \textit{\inlining}, and \textit{\recursive}, respectively corresponding to the leaf node, the root node, and the internal node that the bridge function is in the source FCG.

\begin{figure*}[t]
	\centering
	\subfigure[GCC]{
		\begin{minipage}[t]{0.65\linewidth}
			\centering
			\includegraphics[width=1\textwidth]{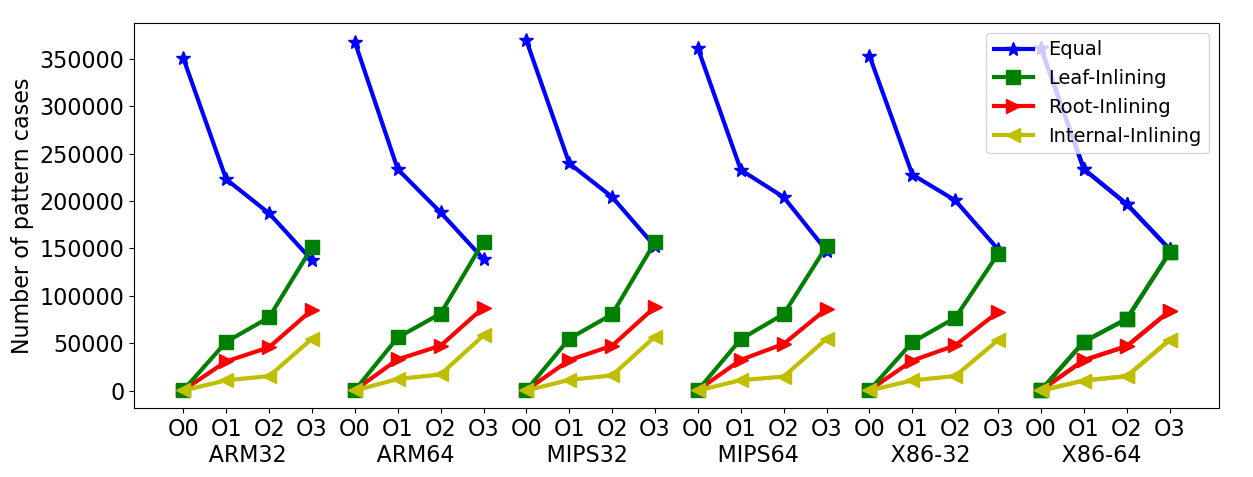}
			\label{fig:pattern_gcc}
			\vspace{-10pt}
		\end{minipage}%
	}%
	
	\vspace{-10pt}
	\subfigure[Clang]{
		\begin{minipage}[t]{0.65\linewidth}
			\centering
			\includegraphics[width=1\textwidth]{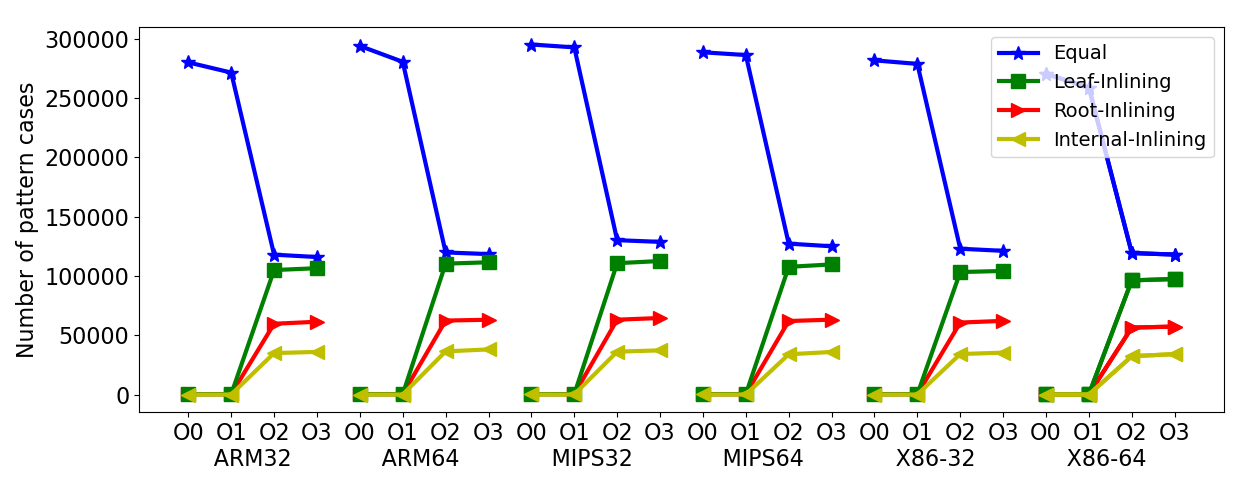}
			\label{fig:pattern_clang}
			\vspace{-10pt}
		\end{minipage}%
	}%
	\vspace{-15pt}
	\caption{Distribution of cross-inlining matching patterns}
	\label{fig:pattern_distribution}
	\vspace{-8pt}
\end{figure*}

As shown in Figure~\ref{fig:matching_pattern_a}, when the bridge function is a leaf node, we classify this cross-inlining pattern as \textit{\inlined}. It is the same as the case in Figure~\ref{fig:motivating_example}. When the bridge function is inlined to the binary function with inlining, its content is surrounded by the source function which calls it and its structure can be mainly reserved.

When the bridge function is a root node as shown in Figure~\ref{fig:matching_pattern_b}, the bridge function will inline other source functions into the body. We classify it as \textit{\inlining}. As a result, though the binary function with inlining still starts with the content of the bridge function, the content of the bridge function is split into several parts and filled in the content of inlined functions.

When the bridge function is an internal node of the source function as shown in Figure~\ref{fig:matching_pattern_c}, the bridge function will not only be inlined but also inline other source functions.  We classify it as \textit{\recursive}. As a result, the function content is split and distributed in the binary functions with inlining.

We further summarize the distribution of these three cross-inlining patterns along with the traditional \textit{Equal} pattern in \textit{Dataset-Inlining} and \textit{Dataset-NoInlining}. Since the statistics of the compilers in the same family present similar distribution~\cite{Tosem}, we aggregate the statistics of the same compiler families. As shown in Figure~\ref{fig:pattern_distribution}, we obtain the distribution of these patterns under 6 architectures, 2 compiler families, and 4 optimizations. 

From Figure~\ref{fig:pattern_distribution}, we first find an increasing trend of cross-inlining patterns from O0 to O3. In detail, we observe that the \textit{\inlined} pattern is more common than \textit{\inlining} and \textit{\recursive} patterns, and even exceeds the number of \textit{Equal} pattern cases when applying O3. The large number of cross-inlining pattern cases indicates the importance of cross-inlining binary function similarity detection.

When comparing the cross-inlining patterns under GCC and Clang, we observe a different distribution. The number of cross-inlining cases keeps increasing from O0 to O3 under GCC, while it only increases from O1 to O2 under Clang. That is because GCC is increasingly adding inlining options to facilitate inlining, while Clang uses \textit{``always inliner''} pass in O0 and O1, and \textit{``inliner''} pass in O2 and O3.

The statistics of cross-inlining patterns under different architectures remain similar. The different architectures do not influence compilers on their inlining decisions.

\subsection{Cross-inlining Matching Evaluation}
\label{sec:evaluaton_of_existing_works}

After analysis of these cross-inlining patterns, we want to evaluate existing works on matching these cross-inlining pairs. Here, we select three existing works, including  Gemini~\cite{Gemini}, Safe~\cite{safe}, and GMN~\cite{GMN_matching}, to evaluate their performance on cross-inlining binary function similarity detection.

\textcolor{revision}{Gemini, Safe, and GMN are state-of-the-art works that have been compared in many research works~\cite{vulseeker, trex, yu2020order, b2b_evaluation}.
Gemini uses a GNN model called Structure2vec~\cite{structure2vec} with manually engineered features to compute function embeddings. SAFE uses the self-attentive sentence encoder~\cite{lin2017structured} to learn cross-architecture function embeddings. GMN proposes a graph matching network with a cross-graph attention-based matching mechanism.}

We use their released models and we run them to calculate the similarity of the cross-inlining pairs. In detail, for these three cross-inlining patterns, we first randomly select 40,000 positive pairs and 40,000 negative pairs from five projects (also testing projects in Section~\ref{sec:experiments}) in \textit{Dataset-Inlining} and \textit{Dataset-NoInlining} respectively. Then we run these models to calculate the pair similarities. Finally, we use Accuracy, Precision, Recall, F1-score, and AUC to evaluate the effectiveness of these models.

As the existing model only produces similarities between function pairs, we can only calculate their AUC before setting the threshold. Then, to calculate other metrics, we tried the threshold from $0.5$ to $0.95$ by the step of $0.05$. When the similarity exceeds the threshold, we regard it as a positive pair, otherwise, a negative pair. Finally, we select the threshold that results in the best F1 and summarize the results as shown in Table~\ref{tab:evaluation_of_existing_works}.

\textcolor{revision}{As shown in Table~\ref{tab:evaluation_of_existing_works}, we use ``Leaf'' as the abbreviation of ``\inlined'', ``Root'' as the abbreviation of ``\inlining'', and ``Internal'' as the abbreviation of ``\recursive''. We also use the same abbreviations in the rest tables.}

\begin{table}[t]
	\caption{Effectiveness of existing works on cross-inlining}
	\vspace{-9pt}
	\scalebox{0.9}{
		\begin{tabular}{c|c|c|c|c|c|c}
			\hline
			Method                  & Pattern   & Accuracy & Precision & Recall & F1   & AUC           \\ \hline
			\multirow{3}{*}{Gemini} & \inlinedshort   & 0.50     & 0.50      & 1.00   & 0.67 & 0.60          \\ \cline{2-7} 
			& \inliningshort  & 0.71     & 0.69      & 0.77   & 0.73 & 0.79 \\ \cline{2-7} 
			& \recursiveshort & 0.73     & 0.70      & 0.80   & 0.75 & 0.57          \\ \hline
			\multirow{3}{*}{Safe}   & \inlinedshort   & 0.53     & 0.53      & 0.53   & 0.53 & 0.54          \\ \cline{2-7} 
			& \inliningshort  & 0.67     & 0.62      & 0.85   & 0.72 & 0.74 \\ \cline{2-7} 
			& \recursiveshort & 0.52     & 0.52      & 0.71   & 0.60 & 0.55          \\ \hline
			\multirow{3}{*}{GMN}    & \inlinedshort   & 0.56     & 0.56      & 0.56   & 0.56 & 0.59          \\ \cline{2-7} 
			& \inliningshort  & 0.74     & 0.71      & 0.80   & 0.75 & 0.83 \\ \cline{2-7} 
			& \recursiveshort & 0.57     & 0.57      & 0.57   & 0.57 & 0.61          \\ \hline
		\end{tabular}
	}
	\label{tab:evaluation_of_existing_works}
	\vspace{-8pt}
\end{table}

In general, compared with their performance on equal pairs~\cite{b2b_evaluation}, existing works suffer a performance loss when detecting cross-inlining pairs. For example, Safe can achieve an AUC of more than 0.90 when detecting equal pairs~\cite{b2b_evaluation}, while it can only achieve an AUC of 0.54 on detecting \textit{\inlined} pairs, 0.74 on \textit{\inlining} pairs, and 0.55 on \textit{\recursive} pairs. In the worst case, the best F1-score of Safe is only 0.53, while regarding all the detection pairs as positive pairs will obtain an F1-score of 0.67.

\begin{figure*}[t]
	\centering
	\includegraphics[width=0.95\textwidth]{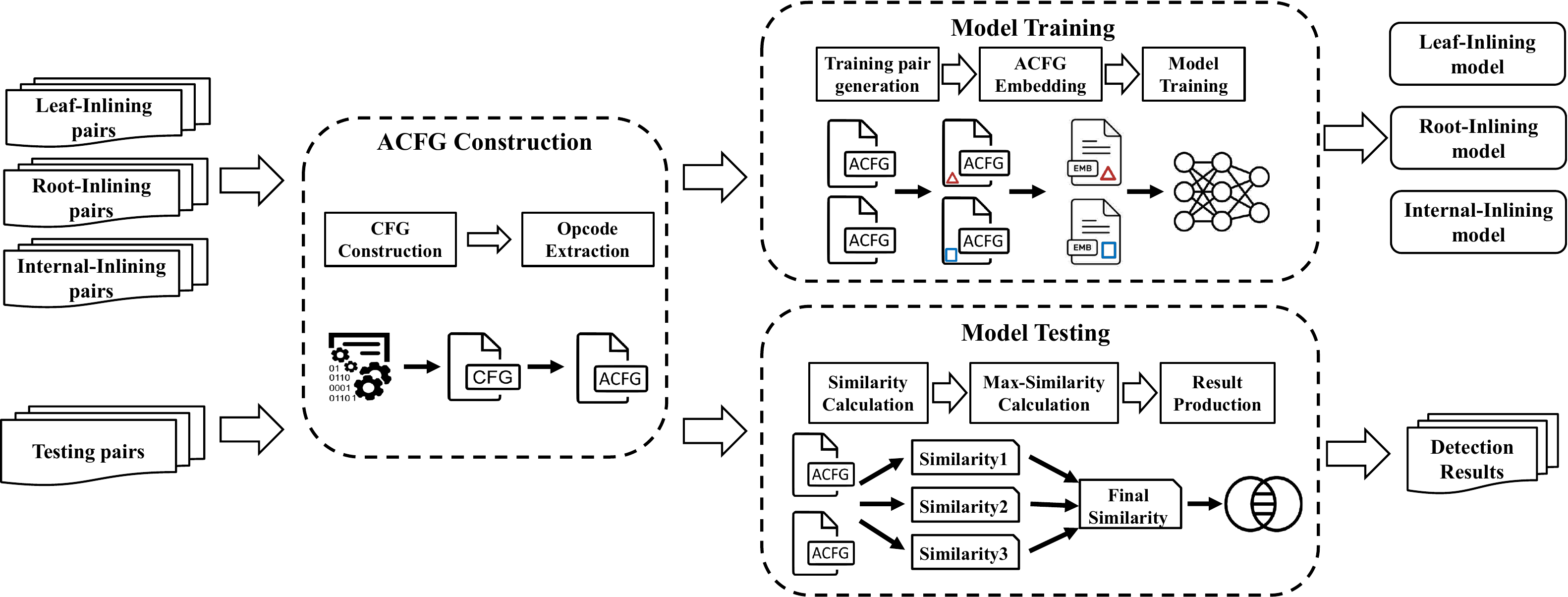}
	\vspace{-5pt}
	\caption{Overview of CI-Detector}
	\label{fig:method_overview}
	\vspace{-10pt}
\end{figure*}

Existing works tend to perform better on \textit{\inlining} pairs compared with \textit{\inlined} and \textit{\recursive} pairs. As shown in Table~\ref{tab:evaluation_of_existing_works}, Gemini can achieve an AUC of 0.79 on \textit{\inlining} pairs while it can only obtain an AUC of 0.60 and 0.57 on \textit{\inlined} and \textit{\recursive} pairs. It can be attributed to inlining conventions. As small functions are more likely to be inlined, the bridge function still plays a major part in the binary function with inlining in the \textit{\inlining} pairs. However, in \textit{\inlined} and \textit{\recursive} pairs, the bridge functions are those inlined functions that account for the small parts, making the detection more challenging.

Besides, GMN performs better on \textit{\inlining} pairs. For example, GMN can achieve an AUC of 0.83 when detecting \textit{\inlining} pairs, where it can obtain an F1-score of 0.75. GMN has also proven to be more effective in normal binary similarity detection~\cite{b2b_evaluation}. In this work, we will use a neural network with a similar architecture to GMN to conduct cross-inlining binary similarity detection.

\section{Method}
\label{sec:method}

In this section, we propose a method, 
named \textbf{CI-Detector} (\textbf{C}ross-\textbf{I}nlining \textbf{D}etector), 
for cross-inlining binary function similarity detection.

Figure~\ref{fig:method_overview} shows the overview of CI-Detector. In general, CI-Detector first extracts ACFGs for these cross-inlining pairs and trains three models for \textit{\inlined}, \textit{\inlining}, and \textit{\recursive} patterns respectively. Then these models are used to calculate similarities for testing pairs and these similarities are aggregated to produce the final similarities.

In this section, we will introduce the ACFG Construction, Model Training, and Model Testing process of CI-Detector.

\subsection{ACFG Construction}

Accurate representation of function semantics is the basis of binary similarity detection. Structural representations, such as CFG, are considered stable representations whose structure varies little for similar code~\cite{binary_similarity_survey}. In this paper, we combine the CFG and the opcodes in the basic blocks to form the ACFG, as the representation of functions.

\begin{figure*}[t]
	\centering
	\subfigure[CFG of \textit{do\_free\_upto}]{
		\begin{minipage}[t]{0.25\linewidth}
			\centering
			\includegraphics[width=0.9\textwidth]{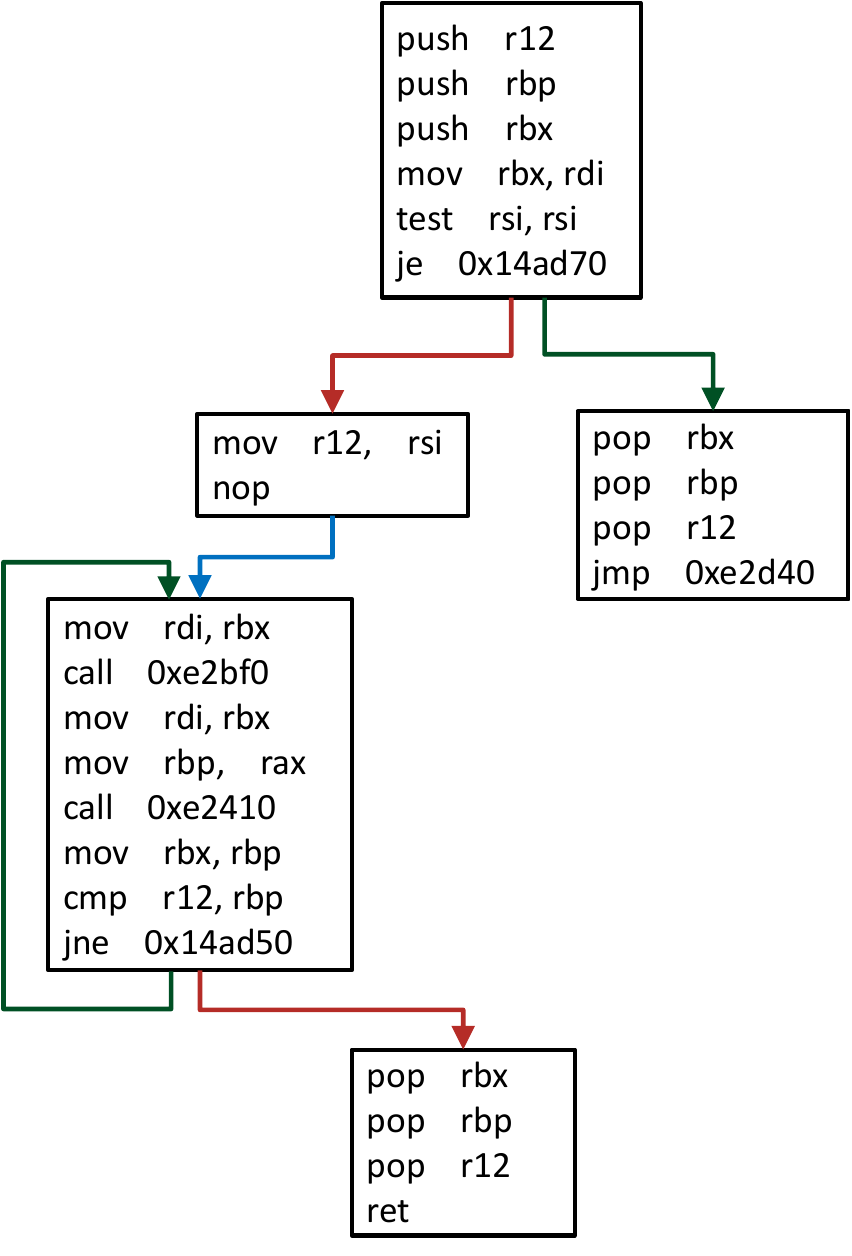}
			\label{fig:acfg_example_1a}
			\vspace{-10pt}
		\end{minipage}%
	}%
	\centering
	\subfigure[ACFG of \textit{do\_free\_upto}]{
		\begin{minipage}[t]{0.25\linewidth}
			\centering
			\includegraphics[width=0.9\textwidth]{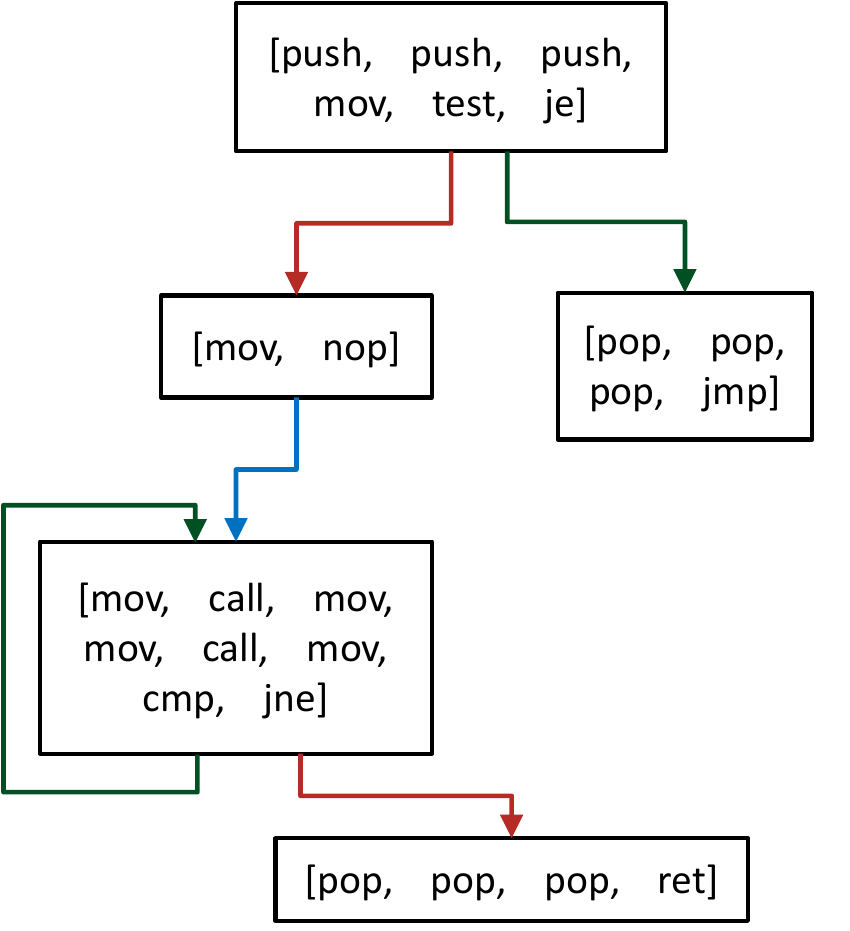}
			\label{fig:acfg_example_1b}
			\vspace{-10pt}
		\end{minipage}%
	}%
	\subfigure[CFG of \textit{CMS\_final}]{
		\begin{minipage}[t]{0.25\linewidth}
			\centering
			\includegraphics[width=0.9\textwidth]{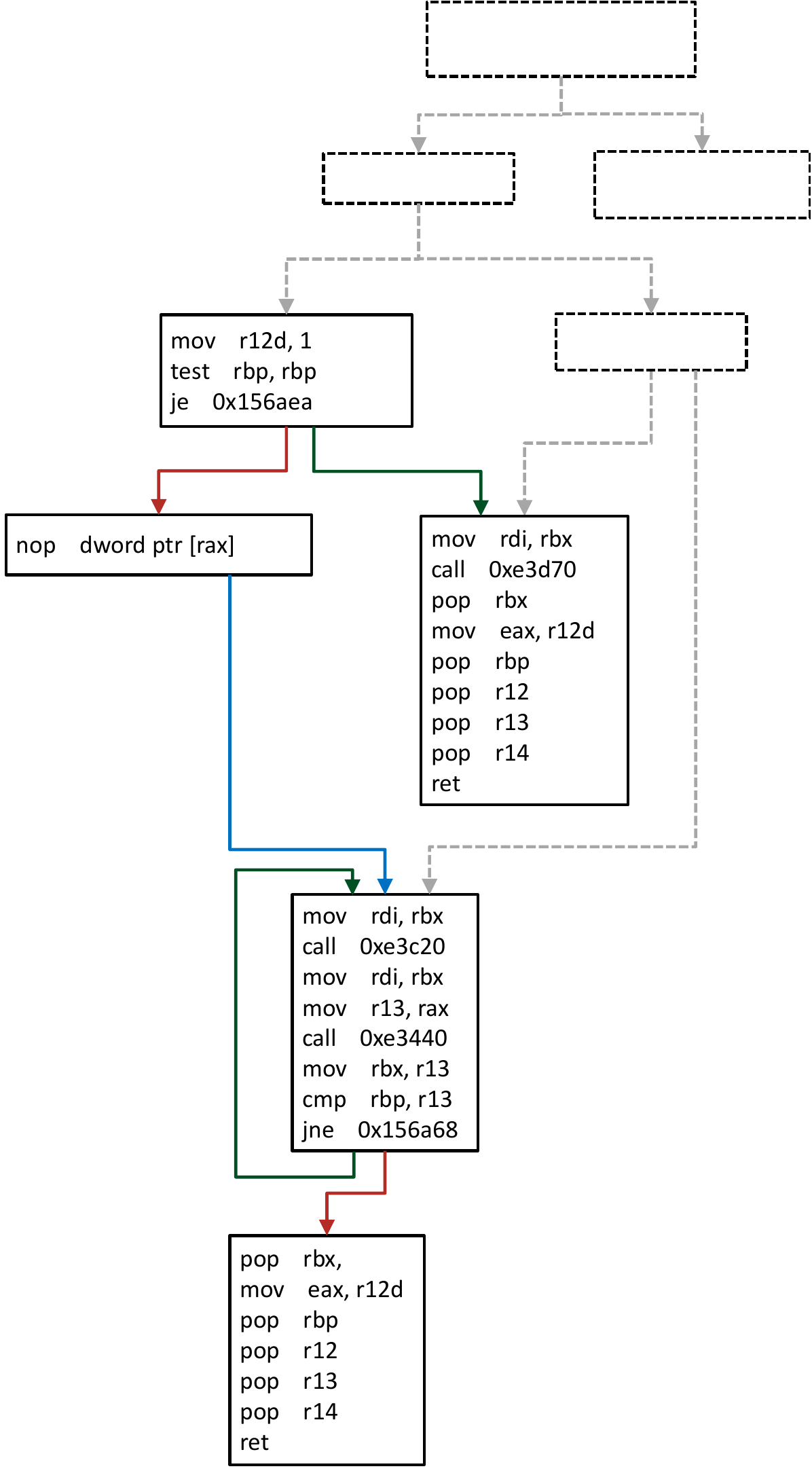}
			\label{fig:acfg_example_2a}
			\vspace{-10pt}
		\end{minipage}%
	}%
	\subfigure[ACFG of \textit{CMS\_final}]{
		\begin{minipage}[t]{0.25\linewidth}
			\centering
			\includegraphics[width=0.9\textwidth]{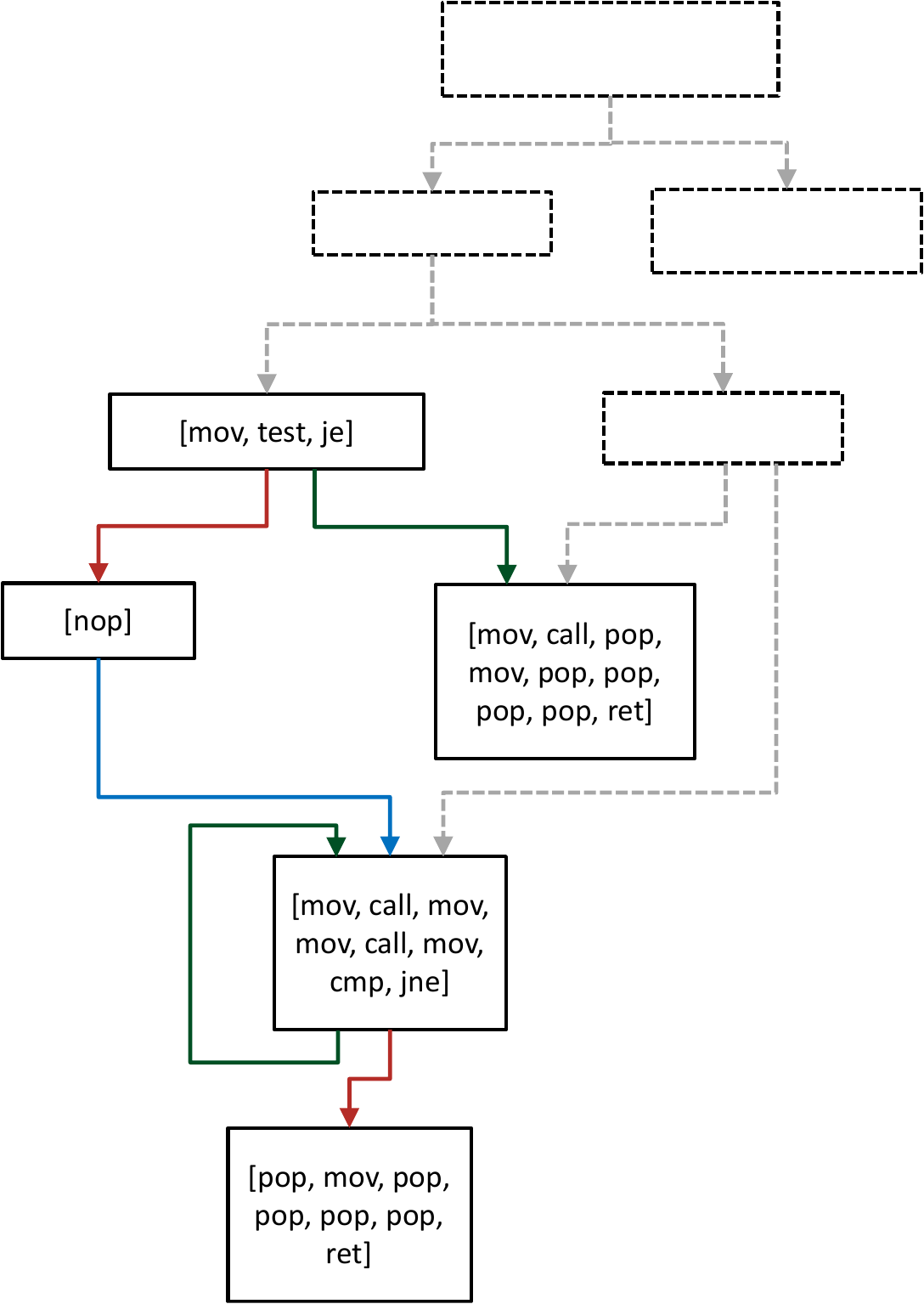}
			\label{fig:acfg_example_2b}
			\vspace{-10pt}
		\end{minipage}%
	}%
	\vspace{-15pt}
	\caption{An example of ACFG construction}
	\label{fig:acfg_example}
	\vspace{-5pt}
\end{figure*}

Figure~\ref{fig:acfg_example} shows an example of constructing ACFG. Figure~\ref{fig:acfg_example_1a} shows the CFG of the function \textit{do\_free\_upto}, we use rectangles to represent basic blocks and arrows to represent control flows. We also present the disassembled instructions in every basic block. An instruction consists of one opcode and zero or more operands. These operands can be registers, immediate values, or addresses.

Figure~\ref{fig:acfg_example_2a} shows the CFG of the function \textit{CMS\_final}. The function \textit{do\_free\_upto} and the function \textit{CMS\_final} is a cross-inlining pair. The dotted rectangles in Figure~\ref{fig:acfg_example_2a} represent binary code from source function \textit{CMS\_final}, and rectangles with solid lines represent binary code from the bridge function \textit{do\_free\_upto}. 

When comparing the binary code in the cross-inlining pair compiled from the same bridge function, we noticed that the operands such as registers and addresses are different in these two functions. As the bridge function shares different contexts in these two binaries, function addresses have been changed and some registers have already been used in the binary function with inlining. Thus, these operands are not robust when conducting cross-inlining similarity detection. Instead, the opcodes in these basic blocks remain similar. Thus, we extract opcodes as attributes of nodes to form the ACFG.

Figure~\ref{fig:acfg_example_1b} and Figure~\ref{fig:acfg_example_2b} are the corresponding ACFGs of function \textit{do\_free\_upto} and function \textit{CMS\_final}. Comparing these two ACFGs, we noticed that the corresponding nodes share similar codes. Even for some nodes that have content from other source functions, such as the root node, they still share similar important opcodes (mov, test, je) that can help neural networks capture the similar semantics of the basic blocks.

\subsection{Model Training}

Considering the difference between these three cross-inlining patterns, we use GNN~\cite{GMN_matching} to train separate models respectively for these patterns. In detail, we train three models including the \textit{\inlined} model, the \textit{\inlining} model, and the \textit{\recursive} model respectively for \textit{\inlined} pairs, the \textit{\inlining} pairs and the \textit{\recursive} pairs.

\textbf{Training pair generation.} In traditional binary code similarity detection, positive pairs are generated by selecting binary functions compiled from the same source function but compiled by different compilers, using different optimizations, to different architectures. These functions have the same function names, thus functions with the same names can be identified as positive pairs, and functions with different names are negative pairs~\cite{b2b_evaluation}.

However, it is more complex to construct positive and negative pairs in cross-inlining binary similarity detection. Functions with different names may not be negative pairs, and functions with the same name are not necessarily positive pairs (for example, equal pairs are not cross-inlining pairs). To construct positive and negative pairs for cross-inlining, we use the source bridge function as the key to construct a dict to arrange the cross-inlining relations.

\begin{lstlisting}[caption={An example of the cross-inlining relation}, label=lst1]
{
  "do_free_upto": {
    "equal": [
      "do_free_upto"
    ],
    "cross-inlining": [
      "CMS_final",
      "CMS_decrypt"
    ]
  }
}
\end{lstlisting}

Listing~\ref{lst1} shows an example of the cross-inlining relations. For the example in Figure~\ref{fig:motivating_example}, source function \textit{do\_free\_upto} is the bridge function of two cross-inlining pairs, including function \textit{CMS\_final} with \textit{do\_free\_upto}, and function \textit{CMS\_decrypt} with \textit{do\_free\_upto}. In Listing~\ref{lst1}, we construct a dict by setting the bridge function as the primary key. Then we set two secondary keys including "equal" and "cross-inlining". In this dict, the binary function \textit{do\_free\_upto} is an "equal" value of the bridge function key \textit{do\_free\_upto} while binary function \textit{CMS\_final} and \textit{CMS\_decrypt} are two "cross-inlining" values. Positive pairs can be generated by randomly selecting one function from the ``equal'' values and one function from "cross-inlining" values under the same source bridge function.

To generate the negative pairs, we randomly select a binary function with inlining that does not appear in the "cross-inlining" values of the current bridge function, and this binary function can be a negative example of the binary functions in the "equal" values. 


\textbf{ACFG Embedding.} To convert ACFG to embeddings, we first represent the node features in vector form. Briefly, we use the bag-of-words model~\cite{bag_of_words} to process these opcodes.

The bag-of-words model converts the opcodes into vectors by counting their frequencies. For example, the opcodes \textit{[push, push, push, mov, test, je]} \textcolor{revision2}{will be expressed as} \textit{\{"push":3, "mov":1, "test":1, "je":1\}}. When we define the key sequence as \textit{[push, mov, test, je]}, then the vector of this node becomes \textit{[3, 1, 1, 1]}.

Then, we convert the full graph into embedding using GNN~\cite{GMN_matching}. The GNN  embedding model comprises 3 parts: (1) an encoder, (2) propagation layers, and (3) an aggregator. 

First, the encoder maps the node features to the initial node vectors through an MLP \textcolor{revision2}{(Multi-layer Perceptron)}. Then, a propagation layer obtains new node representations by aggregating the representations of the node and its neighbors. Through multiple layers of propagation, the representation for each node will accumulate information in its local neighborhood. Finally, the aggregator will aggregate all the node representations to generate final graph representations. Through the above steps, we obtain the embeddings of ACFGs.

\textbf{Model Training.} For each function pair $(G1, G2)$ in the cross-inlining dataset, we can generate embeddings for their ACFGs. Then we use the Euclidean similarity to calculate their similarity and use the following margin-based pairwise loss to optimize the parameters:
\begin{equation}
	Loss = \max \left \{ 0, \gamma - t(1-d(G1, G1) \right \}
	\label{equ:loss}
\end{equation}
where $ t\in \left \{ -1, 1 \right \} $ is the label for this pair, $ \gamma > 0 $ is a margin parameter, and $d(G1, G1)$ is the Euclidean distance. This loss encourage $d(G1, G1) < 1-\gamma$ when the pair is similar ($t = 1$), and $d(G1, G1) > 1+\gamma$ when $t = -1$. Then we use the gradient descent based algorithms to optimize the loss function.

Instead of training a single model on all cross-inlining patterns, we choose to train three models for the three cross-inlining patterns respectively. Considering the difference between each pattern, the model can better fit the characteristics of a single pattern.

After respectively training on the \textit{\inlined} pairs, \textit{\inlining} pairs, and \textit{\recursive} pairs, we obtained three models for these three cross-inlining patterns.

\subsection{Model Testing}
\label{sec:testing}

As shown in Figure~\ref{fig:method_overview}, when inputting the testing pairs, we use the \textit{\inlined} model, \textit{\inlining} model, and \textit{\recursive} model to produce three similarities for each testing pair. The similarity is calculated through the following equation:
\begin{equation}
	sim(G1, G2) = \frac{1}{1+d(G1, G2)} 
	\label{equ:sim}
\end{equation}
Through Equation~\ref{equ:sim}, we normalize the similarity to $(0, 1]$.

Then we use the maximum similarity as the final similarity of the testing pair. The design of similarity calculation is based on the following reasons. On the one hand, when the input pair is a positive pair, it will obtain at least one high similarity. Using the maximum similarity can help identify the positive pairs. On the other hand, when the input pair is a negative pair, these three models will produce low similarities. The maximum similarity will be low which indicates that it is a negative pair. 

After obtaining the similarity, we still need a threshold to distinguish positive pairs and negative pairs. We try different thresholds in the training cross-inlining pairs and find that the threshold $0.55$ produces the best F1 score. Then we use this threshold to compare with similarities of testing pairs and produce the labels for them.

\section{Experiments}
\label{sec:experiments}

In this section, we will first introduce the setup of our experiments. Then, we answer the following research questions to validate the performance of CI-Detector.

\newtheorem{question}{\textbf{RQ}}
\newcounter{myrq}
\setcounter{myrq}{1}
\renewcommand\themyrq{\arabic{myrq}}
\renewcommand\thequestion{\themyrq}

\begin{question}
	\label{RQ-cidetector}
	\textit{Can CI-Detector effectively detect the cross-inlining binary code similarity?}
\end{question}
\addtocounter{myrq}{1}

\begin{question}
	\label{RQ-bingo}
	\textit{How effective is CI-Detector, compared to existing works?}
\end{question}
\addtocounter{myrq}{1}

\begin{question} 
	\label{RQ-ablation}
	\textit{What is the contribution of each model in CI-Detector?}
\end{question}
\addtocounter{myrq}{1}


\begin{question} 
	\label{RQ-efficiency}
	\textit{Can CI-Detector efficiently detect the cross-inlining similarity?}
\end{question}
\addtocounter{myrq}{1}

\subsection{Study Setup}

\subsubsection{Evaluation Dataset.} We select the dataset constructed in Section~\ref{sec:empirical_study} to evaluate CI-Detector. \textcolor{revision}{This dataset is constructed using 9 compilers, 4 optimizations, to 6 architectures, resulting in a total of 216 combinations. It contains several widely used projects, such as coreutils~\cite{coreutils} and binutils~\cite{binutils}.  5,621,140 cross-inlining pairs are identified in this dataset.}

\subsubsection{Experiment Setting.} CI-Detector is trained on cross-inlining pairs in the training dataset and predicts unknown pairs in the testing dataset. To generate the training and testing dataset, we split projects in \textit{Dataset-Inlining} and \textit{Dataset-NoInlining} by 80\%, 10\%, and 10\% to generate the training projects, validation projects, and testing projects. 

\textcolor{revision2}{Though there may be cross-project reuse between the training projects and testing projects~\cite{CrossProjectReuse}, we regard it consistent with the application scenario --- the target binary always uses some existing code in one way or another.}

In testing projects, we randomly generate 40,000 positive pairs and 40,000 negative pairs. These pairs are also used in Section~\ref{sec:evaluaton_of_existing_works} to evaluate existing binary code similarity detection works. We use the same set of testing pairs to facilitate comparison.

The binaries of the dataset are stripped to ensure that CI-Detector will not learn from symbols of the functions in cross-inlining pairs. \textcolor{revision}{We used the default parameters in GMN~\cite{GMN_matching} to construct the GNN, where $\gamma$ is set to 0.1 in Equation~\ref{equ:loss}.} CI-Detector is trained with 128 epochs to produce the cross-inlining models where each epoch consumes 40,000 positive pairs and 40,000 negative pairs randomly generated from training projects.

\subsubsection{Evaluation Metrics.} We use Accuracy, Precision, Recall, F1-score, and AUC to evaluate the effectiveness of CI-Detector. These metrics are widely used to evaluate the effectiveness of methods in classification tasks.

\subsubsection{Implementation of CI-Detector.} We use IDA Pro 7.3~\cite{IDAPro}, Capstone~\cite{capstone}, and NetworkX~\cite{hagberg2008exploring} to construct ACFG for binary functions, and we use TensorFlow 1.14~\cite{tensorflow} to implement GNN model. We refer to this repository~\cite{cisco} to help us with the implementation. The whole procedure is implemented in Python and we run all the experiments on a workstation equipped with Ubuntu 18.04, Intel Xeon Gold 6266C, and 1024GB DDR4 RAM.

\subsection{RQ~\ref{RQ-cidetector}: Effectiveness of CI-Detector} 

In this section, we evaluate the effectiveness of CI-Detector by training it on the training projects and testing it on the testing pairs. \textcolor{revision}{As illustrated in Section~\ref{sec:testing}, we select the threshold 0.55, which results in the best F1-score in the training pairs, to produce labels of testing pairs.}

Table~\ref{tab:evaluation_of_all} shows the evaluation results of CI-Detector on cross-inlining testing pairs. 
As shown in Table~\ref{tab:evaluation_of_all}, CI-Detector performs relatively well on all three cross-inlining patterns. In detail, CI-Detector can achieve an F1-score of 0.88 on \textit{\inlined} pairs, 0.89 on \textit{\inlining} pairs, and 0.88 on \textit{\recursive} pairs. Especially, we noticed that CI-Detector achieves a relatively high recall in detecting cross-inlining pairs. For example, CI-Detector can detect 97\% of \textit{\inlined} pairs, 99\% of \textit{\inlining} pairs, and 96\% of \textit{\recursive} pairs. Meanwhile, CI-Detector can achieve an average accuracy of 81\% in detecting these cross-inlining pairs.

\begin{table}[h]
	\caption{Effectiveness of CI-Detector and existing works}
	\vspace{-8pt}
	\scalebox{0.9}{
		\begin{tabular}{c|c|c|c|c|c|c}
			\hline
			Method                       & Pattern  & Accuracy & Precision & Recall & F1   & AUC  \\ \hline
			\multirow{3}{*}{Gemini}      & Leaf     & 0.50     & 0.50      & 1.00   & 0.67 & 0.60 \\ \cline{2-7} 
			& Root     & 0.71     & 0.69      & 0.77   & 0.73 & 0.79 \\ \cline{2-7} 
			& Internal & 0.73     & 0.70      & 0.80   & 0.75 & 0.57 \\ \hline
			\multirow{3}{*}{Safe}        & Leaf     & 0.53     & 0.53      & 0.53   & 0.53 & 0.54 \\ \cline{2-7} 
			& Root     & 0.67     & 0.62      & 0.85   & 0.72 & 0.74 \\ \cline{2-7} 
			& Internal & 0.52     & 0.52      & 0.71   & 0.60 & 0.55 \\ \hline
			\multirow{3}{*}{GMN}         & Leaf     & 0.56     & 0.56      & 0.56   & 0.56 & 0.59 \\ \cline{2-7} 
			& Root     & 0.74     & 0.71      & 0.80   & 0.75 & 0.83 \\ \cline{2-7} 
			& Internal & 0.57     & 0.57      & 0.57   & 0.57 & 0.61 \\ \hline
			\multirow{3}{*}{Bingo}       & Leaf     & 0.71     & 0.67      & 0.82   & 0.74 & 0.73 \\ \cline{2-7} 
			& Root     & 0.75     & 0.68      & 0.95   & 0.79 & 0.80 \\ \cline{2-7} 
			& Internal & 0.69     & 0.66      & 0.79   & 0.72 & 0.72 \\ \hline
			\multirow{3}{*}{Asm2Vec}     & Leaf     & 0.50     & 0.50      & 1.00   & 0.67 & 0.54 \\ \cline{2-7} 
			& Root     & 0.50     & 0.50      & 1.00   & 0.67 & 0.55 \\ \cline{2-7} 
			& Internal & 0.50     & 0.50      & 1.00   & 0.67 & 0.54 \\ \hline
			\multirow{3}{*}{CI-Detector} & Leaf     & 0.87     & 0.81      & 0.97   & 0.88 & 0.89 \\ \cline{2-7} 
			& Root     & 0.88     & 0.81      & 0.99   & 0.89 & 0.87 \\ \cline{2-7} 
			& Internal & 0.87     & 0.81      & 0.96   & 0.88 & 0.88 \\ \hline
		\end{tabular}
		\label{tab:evaluation_of_all}
		\vspace{-5pt}
	}
\end{table}

\textcolor{revision}{We then respectively analyze the false negatives and false positives of CI-Detector. False negatives are the cases where the testing pair is positive while CI-Detector classifies it as negative. False positives are the cases where the testing pair is negative while CI-Detector classifies it as positive. As CI-Detector obtains a high recall, the false negatives rate is only 3\%. }

\textcolor{revision}{The false negatives are mainly due to the huge difference in function size of some positive cross-inlining pairs.
	One false negative example is the function \textit{PUSH\_CODE} in \textit{Dataset-NoInlining} with the function \textit{r\_interpret} in \textit{Dataset-Inlining}. Both functions are from the project gawk~\cite{gawk} compiled by gcc-5.5.0 with O3 to X86\_64. We notice that this function pair has a huge difference in their function size. The function \textit{PUSH\_CODE} is only composed of 6 basic blocks with less than 30 instructions, while the function \textit{r\_interpret} is composed of 857 basic blocks with 3,651 instructions. 
	The huge difference not only makes it difficult to learn the cross-inlining relation between these two functions, but also makes it harder to find the location of the inlined function.
} 

\textcolor{revision}{The false positives are mainly due to some small functions with a single functionality. For example, \textit{free\_token} is a function that frees the token sent to this function. However, many binary functions have the action of freeing some variables. They have similar actions with the function \textit{free\_token} but they do not inline such a function. This makes CI-Detector wrongly recognize such a function pair as a cross-inlining pair.}

\begin{tcolorbox}
	
	\textbf{Answering RQ~\ref{RQ-cidetector}:} 
	CI-Detector can effectively detect the cross-inlining binary code similarity. 
	CI-Detector can achieve a precision of 81\% and a recall of 97\% when detecting cross-inlining pairs. 
\end{tcolorbox}

\subsection{RQ~\ref{RQ-bingo}: Compared with Existing Works}

In this section, we compare the performance of CI-Detector with existing works. Apart from the works evaluated in Section~\ref{sec:evaluaton_of_existing_works},  
we also compare CI-Detector with Bingo~\cite{bingo} and Asm2Vec~\cite{asm2vec}, which proposes strategies to simulate function inlining. Bingo and Asm2Vec use manual design rules to search for callee functions that should be inlined and then compare the binary functions after inlining to obtain the similarity. 
\textcolor{revision}{
	We use their released code and model for testing.
}  
Table~\ref{tab:evaluation_of_all} shows the performance of CI-Detector and existing works.

In general, CI-Detector exceeds all state-of-the-art works in terms of all metrics. Compared with existing works, CI-Detector obtained a 10\%-30\% improvement in precision, with a recall of more than 95\%. 
Existing works often suffer from detecting \textit{\inlined} and \textit{\recursive} pairs. However, CI-Detector can perform relatively well on all three inlining patterns.

The inlining strategies can help Bingo detect the \textit{\inlining} pairs with a recall of 95\%, which exceeds all other existing works. However, its strategies only work for the \textit{\inlining} pairs. Its performance on \textit{\inlined} and \textit{\recursive} pairs still suffers from low precision and recall. \textcolor{revision}{That is because the strategies of bingo can only help inline callee functions into query functions and target functions. When the query function is an inlined function, it cannot find the caller functions in the reverse direction.} Besides, Asm2Vec does not perform well on the cross-inlining dataset. Even with the strategies of Bingo, it cannot accurately distinguish the cross-inlining pairs.

CI-Detector, which has respectively trained models for all cross-inlining patterns, can detect \textit{\inlining}, \textit{\inlined}, and \textit{\recursive} pairs all with high coverage. Compared to the inlining strategies of Bingo, CI-Detector does not need the context of query functions and target functions. CI-Detector is more scalable for cross-inlining binary function similarity detection.

\begin{tcolorbox}
	
	\textbf{Answering RQ~\ref{RQ-bingo}:}  
	CI-Detector exceeds all state-of-the-art binary similarity detection works. Although Bingo has proposed strategies for \textit{\inlining} pairs, CI-Detector not only can detect the \textit{\inlining} pairs but also is effective in detecting \textit{\inlined} and \textit{\recursive} pairs. 
\end{tcolorbox}

\subsection{RQ~\ref{RQ-ablation}: Contribution of Each Model} 

To evaluate the contribution of models in CI-Detector, we first create several variants of CI-Detector, including \textit{\inlined}, \textit{\inlining}, \textit{\recursive}, and \textit{Mixed} model. \textit{\inlined}, \textit{\inlining}, and \textit{\recursive} model are methods that use the model trained only on one cross-inlining pattern. \textcolor{revision}{\textit{Mixed} model is directly trained on all the cross-inlining patterns.} Table~\ref{tab:ablation} shows the evaluation results of them.

\begin{table}[h]
	\caption{Effectiveness of CI-Detectors and its variants}
	\vspace{-8pt}
	\centering
	\scalebox{0.9}{
		\begin{tabular}{c|c|c|c|c|c|c}
			\hline
			Method                       & Pattern   & Accuracy & Precision & Recall & F1   & AUC  \\ \hline
			\multirow{3}{*}{\inlinedshort}     & \inlinedshort   & 0.86     & 0.80      & 0.95   & 0.87 & 0.87 \\ \cline{2-7} 
			& \inliningshort  & 0.88     & 0.82      & 0.98   & 0.89 & 0.93 \\ \cline{2-7} 
			& \recursiveshort & 0.86     & 0.81      & 0.94   & 0.87 & 0.87 \\ \hline
			\multirow{3}{*}{\inliningshort}    & \inlinedshort   & 0.79     & 0.80      & 0.79   & 0.79 & 0.85 \\ \cline{2-7} 
			& \inliningshort  & 0.89     & 0.84      & 0.97   & 0.90 & 0.96 \\ \cline{2-7} 
			& \recursiveshort & 0.82     & 0.82      & 0.83   & 0.82 & 0.87 \\ \hline
			\multirow{3}{*}{\recursiveshort}   & \inlinedshort   & 0.85     & 0.80      & 0.92   & 0.86 & 0.87 \\ \cline{2-7} 
			& \inliningshort  & 0.88     & 0.81      & 0.98   & 0.89 & 0.94 \\ \cline{2-7} 
			& \recursiveshort & 0.86     & 0.81      & 0.94   & 0.87 & 0.88 \\ \hline
			\multirow{3}{*}{Mixed}       & \inlinedshort   & 0.84     & 0.80      & 0.89   & 0.85 & 0.86 \\ \cline{2-7} 
			& \inliningshort  & 0.87     & 0.81      & 0.96   & 0.88 & 0.92 \\ \cline{2-7} 
			& \recursiveshort & 0.84     & 0.81      & 0.90   & 0.85 & 0.85 \\ \hline
			\multirow{3}{*}{CI-Detector} & \inlinedshort   & 0.87     & 0.81      & 0.97   & 0.88 & 0.89 \\ \cline{2-7} 
			& \inliningshort  & 0.88     & 0.81      & 0.99   & 0.89 & 0.87 \\ \cline{2-7} 
			& \recursiveshort & 0.87     & 0.81      & 0.96   & 0.88 & 0.88 \\ \hline
	\end{tabular}}
	\label{tab:ablation}
\end{table}

As shown in Table~\ref{tab:ablation}, single models including \textit{\inlined}, \textit{\inlining}, and \textit{\recursive} models have interesting relations among them. For example, the \textit{\inlining} model performs extremely well on the \textit{\inlining} pairs but suffers a loss in recall of detecting \textit{\inlined} and \textit{\recursive} pairs. That indicates that the \textit{\inlining} pattern cannot scale to the other two patterns. However, the \textit{\inlined} and \textit{\recursive} models, which are trained respectively on  \textit{\inlined} and \textit{\recursive} pairs, perform better on the \textit{\inlining} pairs than \textit{\inlined} and \textit{\recursive} pairs. That indicates that the \textit{\inlining} pattern is easier to be learned compared with \textit{\inlined} and \textit{\recursive} patterns.

The result of the \textit{Mixed} model also indicates this relation. When training on all three cross-inlining patterns, we find that the \textit{Mixed} model tends to perform better on the \textit{\inlining} pairs than the \textit{\inlined} and \textit{\recursive} pairs. Especially, \textit{Mixed} model suffer a loss in its recall of detecting all cross-inlining pairs, indicating that mixing these three inlining pattern makes the learning of cross-inlining pattern more difficult.

\begin{figure}[h]
	\centering
	\vspace{-5pt}
	\includegraphics[width=0.4\textwidth]{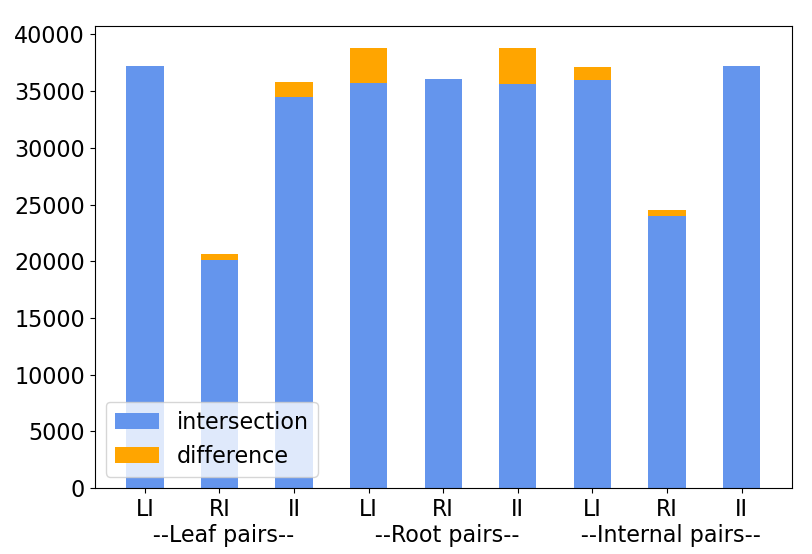}
	\vspace{-10pt}
	\caption{Relation of three cross-inlining models in retrieving positive pairs}
	\label{fig:model_relation}
	\vspace{-7pt}
\end{figure}

CI-Detector improves the performance of detecting all three inlining patterns by combining all three cross-inlining models. Figure~\ref{fig:model_relation} presents the intersection and difference of the identified positive pairs detected by \textit{\inlined} (LI), \textit{\inlining} (RI), and \textit{\recursive} (II) models. For example, the ``RI'' above the ``--Leaf pairs'' means the results of using \textit{\inlining} model to detect \textit{\inlined} pairs. The blue bar represents the intersection of positive pairs identified by \textit{\inlining} model and \textit{\inlined} model, and the orange bar represents the difference of \textit{\inlining} model to \textit{\inlined} model. 

As shown in Figure~\ref{fig:model_relation}, one model can help another model recover hidden positive pairs. For example, \textit{\inlined} and \textit{\recursive} models help \textit{\inlining} model recover thousands of positive \textit{\inlining} pairs. Besides, \textit{\recursive} model can help \textit{\inlined} model recover \textit{\inlined} positive pairs, and \textit{\inlined} model can help \textit{\recursive} model recover \textit{\recursive} positive pairs. As a result, CI-Detector can achieve a recall of 97\% on detecting \textit{\inlined} pairs, 99\% on detecting \textit{\inlining} pairs, and 96\% on detecting \textit{\recursive} pairs.

%
%

\begin{tcolorbox}
	
	\textbf{Answering RQ~\ref{RQ-ablation}:}  
	The \textit{\inlined}, \textit{\inlining}, and \textit{\recursive} models can help each other to recover more positive cross-inlining pairs, which helps CI-Detector to achieve high recall when detecting all three kinds of cross-inlining pairs.
\end{tcolorbox}


\subsection{RQ~\ref{RQ-efficiency}: Efficiency of CI-Detector} 

In this section, we will evaluate the efficiency of CI-Detector. We will first present the training time of CI-Detector. Then we compare the testing time of CI-Detector with existing works.

\begin{figure}[h]
	\centering
	\vspace{-5pt}
	\includegraphics[width=0.4\textwidth]{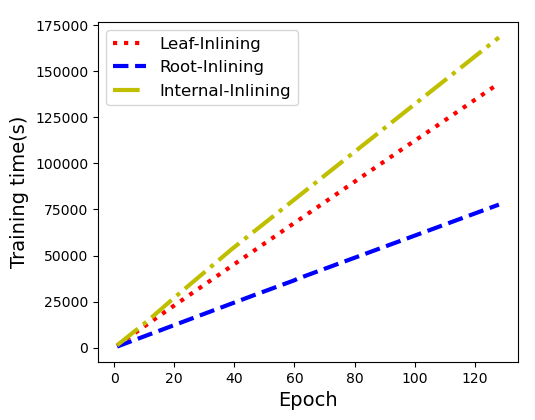}
	\vspace{-10pt}
	\caption{Training time of CI-Detector}
	\label{fig:training_time}
	\vspace{-7pt}
\end{figure}

As CI-Detector contains three models training respectively for three cross-inlining patterns.
We record the training time of these three cross-inlining models as shown in Figure~\ref{fig:training_time}. In general, training on \textit{\inlining} pairs in 128 epochs costs about 20 hours, training on \textit{\inlined} pairs costs about 37 hours, and training on \textit{\recursive} pairs costs about 48 hours. 

Considering that CI-Detector needs to be trained only once, the training time is acceptable.

\begin{figure}[h]
	\centering
	\vspace{-5pt}
	\includegraphics[width=0.4\textwidth]{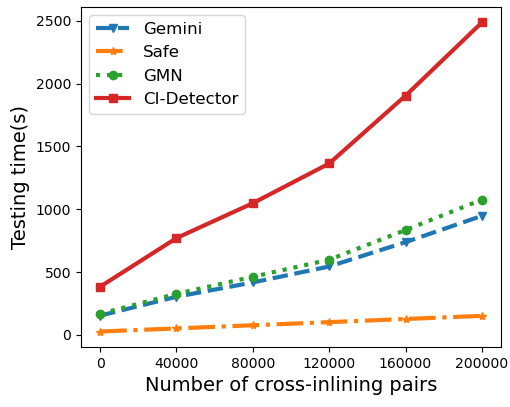}
	\vspace{-8pt}
	\caption{Testing time of CI-Detector}
	\label{fig:testing_time}
	\vspace{-7pt}
\end{figure}

Figure~\ref{fig:testing_time} shows the testing time of Gemini, Safe, GMN, and CI-Detector. Bingo and Asm2vec need to perform the program-level comparison to calculate the similarity between two functions, thus we exclude them. As shown in Figure~\ref{fig:testing_time}, Safe spends less than 0.001s to process a testing pair as it only uses a self-attentive neural network. The GNN-based methods including Gemini and GMN spend 0.005s to process a testing pair on average. CI-Detector, which uses three GNN models, spends about 0.013s to process a testing pair. 

Though CI-Detector spends more time than existing works, it is still efficient as it only needs about 2/3 hours to process 200,000 testing pairs. 

\begin{tcolorbox}
	
	\textbf{Answering RQ~\ref{RQ-efficiency}:}  
	CI-Detector can efficiently detect the cross-inlining similarity. On average, the models in CI-Detector only need 0.013s to produce the similarities of a cross-inlining pair.
\end{tcolorbox}



\section{Threats to validity}
\label{sec:threats}

In this section, we will discuss the threats to validity.

\textbf{Threats to internal validity.} The threat to internal validity is that we use IDA Pro to construct the ACFGs of binary functions. As pointed out by study~\cite{pang2021sok}, accurate disassembling and binary analysis is not easy. The errors in disassembling may affect the ACFG construction and further affect the results.

\textbf{Threats to external validity.} The threat to external validity is the diversity of compilation settings. In this paper, though we have compiled the project with 9 compilers, and 4 optimizations, to 6 architectures, there may be other compilation settings such as non-default optimizations~\cite{10.1145/3453483.3454035}. The inlining patterns learned by our model may not scale to projects with different inlining conventions. However, our method can be easily retrained using other datasets, as long as these datasets can be labeled with inlining labels.

\textbf{Threats to construct validity.} The threat to the construct validity is that our labeling method relies on some third-party tools such as IDA Pro and Dwarf debug tools. As also mentioned in~\cite{Tosem}, these tools can be inaccurate, thus further affecting the identification of function inlining. To help mitigate the influence, we have investigated enormous tools and chosen the most reliable ones. For example, IDA Pro is a state-of-the-art commercial tool and has been widely used in academia~\cite{deepbindiff, pang2021sok} and industry~\cite{IDA_case1, IDA_case2}. 

\section{Conclusion}
\label{sec:conclusion}

In this paper, we propose a pattern-based model named CI-Detector for cross-inlining matching. CI-Detector uses the ACFG to represent the semantics of binary functions and GNN to embed binary functions into vectors. We summarize three cross-inlining patterns in the cross-inlining dataset and CI-Detector respectively trains a model for these three cross-inlining patterns. The testing pairs are input to these three models and all the produced similarities are aggregated to produce the final similarity. We conduct several experiments to evaluate CI-Detector. Results show that CI-Detector can detect cross-inlining pairs with a precision of 81\% and a recall of 97\%, which exceeds all state-of-the-art works.

\begin{acks}
	This work was supported by National Key R\&D Program of China (2022YFB2703500), National Natural Science Foundation of China (62232014, 62272377, 62293501, 62293502, 72241433, 61721002,\\ 62032010, 62002280, 62372368, 62372367), CCF-AFSG Research Fund, China Postdoctoral Science Foundation (2020M683507, 2019TQ0251, 2020M673439), and Young Talent Fund of Association for Science and Technology in Shaanxi, China.
\end{acks}

\bibliographystyle{ACM-Reference-Format}
\bibliography{reference}


\begin{thebibliography}{84}


\ifx \showCODEN    \undefined \def \showCODEN     #1{\unskip}     \fi
\ifx \showDOI      \undefined \def \showDOI       #1{#1}\fi
\ifx \showISBNx    \undefined \def \showISBNx     #1{\unskip}     \fi
\ifx \showISBNxiii \undefined \def \showISBNxiii  #1{\unskip}     \fi
\ifx \showISSN     \undefined \def \showISSN      #1{\unskip}     \fi
\ifx \showLCCN     \undefined \def \showLCCN      #1{\unskip}     \fi
\ifx \shownote     \undefined \def \shownote      #1{#1}          \fi
\ifx \showarticletitle \undefined \def \showarticletitle #1{#1}   \fi
\ifx \showURL      \undefined \def \showURL       {\relax}        \fi
\providecommand\bibfield[2]{#2}
\providecommand\bibinfo[2]{#2}
\providecommand\natexlab[1]{#1}
\providecommand\showeprint[2][]{arXiv:#2}

\bibitem[IDA(2020a)]%
        {IDA_case1}
 \bibinfo{year}{2020}\natexlab{a}.
\newblock \bibinfo{title}{{Reveiws 1 - SciTools}}.
\newblock
  \bibinfo{howpublished}{\url{https://news.sophos.com/en-us/2020/04/26/asnarok/}}.
\newblock
\newblock
\shownote{[Online; accessed 3-September-2021]}.


\bibitem[IDA(2020b)]%
        {IDA_case2}
 \bibinfo{year}{2020}\natexlab{b}.
\newblock \bibinfo{title}{{What’s up, Emotet? CERT Polska}}.
\newblock
  \bibinfo{howpublished}{\url{https://cert.pl/en/posts/2020/02/whats-up-emotet/}}.
\newblock
\newblock
\shownote{[Online; accessed 3-September-2021]}.


\bibitem[gnu(2021)]%
        {gnulib}
 \bibinfo{year}{2021}\natexlab{}.
\newblock \bibinfo{title}{{Gnulib - The GNU Portability Library}}.
\newblock \bibinfo{howpublished}{\url{https://www.gnu.org/software/gnulib/}}.
\newblock
\newblock
\shownote{[Online; accessed 23-April-2022]}.


\bibitem[IDA(2021)]%
        {IDAPro}
 \bibinfo{year}{2021}\natexlab{}.
\newblock \bibinfo{title}{{IDA Pro Disassembler and Debugger - Hex Rays}}.
\newblock \bibinfo{howpublished}{\url{https://www.hex-rays.com/ida-pro/}}.
\newblock
\newblock
\shownote{[Online; accessed 20-April-2022]}.


\bibitem[wik(2022)]%
        {wiki_inlining}
 \bibinfo{year}{2022}\natexlab{}.
\newblock \bibinfo{title}{{Inline expansion}}.
\newblock
  \bibinfo{howpublished}{\url{https://en.wikipedia.org/wiki/Inline_expansion}}.
\newblock
\newblock
\shownote{[Online; accessed 23-April-2022]}.


\bibitem[Cod(2022)]%
        {CodeResue}
 \bibinfo{year}{2022}\natexlab{}.
\newblock \bibinfo{title}{{Synopsys 2022 open source security and risk analysis
  report}}.
\newblock
  \bibinfo{howpublished}{\url{https://www.synopsys.com/software-integrity/resources/analyst-reports/open-source-security-risk-analysis.html}}.
\newblock
\newblock
\shownote{[Online; accessed 29-May-2023]}.


\bibitem[TPL(2022)]%
        {TPL}
 \bibinfo{year}{2022}\natexlab{}.
\newblock \bibinfo{title}{{VDC Research White Paper | Finding Sources of
  Security in the Complex Software Supply Chains of Tomorrow}}.
\newblock
  \bibinfo{howpublished}{\url{https://codesonar.grammatech.com/wp-form-vdc-research-software-supply-chain}}.
\newblock
\newblock
\shownote{[Online; accessed 29-May-2023]}.


\bibitem[bin(2023)]%
        {binutils}
 \bibinfo{year}{2023}\natexlab{}.
\newblock \bibinfo{title}{{Binutils - GNU Porject}}.
\newblock \bibinfo{howpublished}{\url{https://www.gnu.org/software/binutils/}}.
\newblock
\newblock
\shownote{[Online; accessed 25-July-2023]}.


\bibitem[cap(2023)]%
        {capstone}
 \bibinfo{year}{2023}\natexlab{}.
\newblock \bibinfo{title}{{capstone-PyPI}}.
\newblock \bibinfo{howpublished}{\url{https://pypi.org/project/capstone/}}.
\newblock
\newblock
\shownote{[Online; accessed 13-July-2022]}.


\bibitem[cis(2023)]%
        {cisco}
 \bibinfo{year}{2023}\natexlab{}.
\newblock \bibinfo{title}{{Cisco-Talos/binary-function-similarity}}.
\newblock
  \bibinfo{howpublished}{\url{https://github.com/Cisco-Talos/binary_function_similarity}}.
\newblock
\newblock
\shownote{[Online; accessed 13-July-2022]}.


\bibitem[cor(2023)]%
        {coreutils}
 \bibinfo{year}{2023}\natexlab{}.
\newblock \bibinfo{title}{{Coreutils - GNU core utilities}}.
\newblock
  \bibinfo{howpublished}{\url{https://www.gnu.org/software/coreutils/}}.
\newblock
\newblock
\shownote{[Online; accessed 25-July-2023]}.


\bibitem[gaw(2023)]%
        {gawk}
 \bibinfo{year}{2023}\natexlab{}.
\newblock \bibinfo{title}{{Gawk - GNU Project}}.
\newblock \bibinfo{howpublished}{\url{https://www.gnu.org/software/gawk/}}.
\newblock
\newblock
\shownote{[Online; accessed 25-July-2023]}.


\bibitem[CVE(2023a)]%
        {CVE-2014-3569}
 \bibinfo{year}{2023}\natexlab{a}.
\newblock \bibinfo{title}{{NVD - CVE-2014-3569}}.
\newblock
  \bibinfo{howpublished}{\url{https://nvd.nist.gov/vuln/detail/cve-2014-3569}}.
\newblock
\newblock
\shownote{[Online; accessed 6-June-2023]}.


\bibitem[CVE(2023b)]%
        {CVE-2015-1792}
 \bibinfo{year}{2023}\natexlab{b}.
\newblock \bibinfo{title}{{NVD - CVE-2015-1792}}.
\newblock
  \bibinfo{howpublished}{\url{https://nvd.nist.gov/vuln/detail/CVE-2015-1792}}.
\newblock
\newblock
\shownote{[Online; accessed 29-May-2023]}.


\bibitem[ope(2023)]%
        {openssl}
 \bibinfo{year}{2023}\natexlab{}.
\newblock \bibinfo{title}{OpenSSL}.
\newblock \bibinfo{howpublished}{\url{https://www.openssl.org/}}.
\newblock


\bibitem[git(2023)]%
        {github_repo}
 \bibinfo{year}{2023}\natexlab{}.
\newblock \bibinfo{title}{{Source code and dataset. }}.
\newblock
  \bibinfo{howpublished}{\url{https://github.com/island255/cross-inlining_binary_function_similarity}}.
\newblock
\newblock
\shownote{[Online; accessed 21-July-2023]}.


\bibitem[ten(2023)]%
        {tensorflow}
 \bibinfo{year}{2023}\natexlab{}.
\newblock \bibinfo{title}{{tensorflow-PyPI}}.
\newblock \bibinfo{howpublished}{\url{https://pypi.org/project/tensorflow/}}.
\newblock
\newblock
\shownote{[Online; accessed 13-July-2022]}.


\bibitem[Andersson(2009)]%
        {andersson2009evaluation}
\bibfield{author}{\bibinfo{person}{P{\"a}r Andersson}.}
  \bibinfo{year}{2009}\natexlab{}.
\newblock \bibinfo{title}{Evaluation of inlining heuristics in industrial
  strength compilers for embedded systems}.
\newblock
\newblock


\bibitem[Ban et~al\mbox{.}(2021)]%
        {ban2021b2smatcher}
\bibfield{author}{\bibinfo{person}{Gu Ban}, \bibinfo{person}{Lili Xu},
  \bibinfo{person}{Yang Xiao}, \bibinfo{person}{Xinhua Li},
  \bibinfo{person}{Zimu Yuan}, {and} \bibinfo{person}{Wei Huo}.}
  \bibinfo{year}{2021}\natexlab{}.
\newblock \showarticletitle{B2SMatcher: fine-Grained version identification of
  open-Source software in binary files}.
\newblock \bibinfo{journal}{\emph{Cybersecurity}} \bibinfo{volume}{4},
  \bibinfo{number}{1} (\bibinfo{year}{2021}), \bibinfo{pages}{1--21}.
\newblock


\bibitem[Chandramohan et~al\mbox{.}(2016)]%
        {bingo}
\bibfield{author}{\bibinfo{person}{Mahinthan Chandramohan},
  \bibinfo{person}{Yinxing Xue}, \bibinfo{person}{Zhengzi Xu},
  \bibinfo{person}{Yang Liu}, \bibinfo{person}{Chia~Yuan Cho}, {and}
  \bibinfo{person}{Hee Beng~Kuan Tan}.} \bibinfo{year}{2016}\natexlab{}.
\newblock \showarticletitle{Bingo: Cross-architecture cross-os binary search}.
  In \bibinfo{booktitle}{\emph{Proceedings of the 2016 24th ACM SIGSOFT
  International Symposium on Foundations of Software Engineering}}.
  \bibinfo{pages}{678--689}.
\newblock


\bibitem[Cooper et~al\mbox{.}(2008)]%
        {DBLP:conf/cc/CooperHW08}
\bibfield{author}{\bibinfo{person}{Keith~D. Cooper},
  \bibinfo{person}{Timothy~J. Harvey}, {and} \bibinfo{person}{Todd Waterman}.}
  \bibinfo{year}{2008}\natexlab{}.
\newblock \showarticletitle{An Adaptive Strategy for Inline Substitution}. In
  \bibinfo{booktitle}{\emph{Compiler Construction, 17th International
  Conference, {CC} 2008, Held as Part of the Joint European Conferences on
  Theory and Practice of Software, {ETAPS} 2008, Budapest, Hungary, March 29 -
  April 6, 2008. Proceedings}} \emph{(\bibinfo{series}{Lecture Notes in
  Computer Science}, Vol.~\bibinfo{volume}{4959})}.
  \bibinfo{publisher}{Springer}, \bibinfo{pages}{69--84}.
\newblock
\urldef\tempurl%
\url{https://doi.org/10.1007/978-3-540-78791-4\_5}
\showDOI{\tempurl}


\bibitem[Dai et~al\mbox{.}(2016)]%
        {structure2vec}
\bibfield{author}{\bibinfo{person}{Hanjun Dai}, \bibinfo{person}{Bo Dai}, {and}
  \bibinfo{person}{Le Song}.} \bibinfo{year}{2016}\natexlab{}.
\newblock \showarticletitle{Discriminative embeddings of latent variable models
  for structured data}. In \bibinfo{booktitle}{\emph{International conference
  on machine learning}}. PMLR, \bibinfo{pages}{2702--2711}.
\newblock


\bibitem[David et~al\mbox{.}(2016)]%
        {Esh}
\bibfield{author}{\bibinfo{person}{Yaniv David}, \bibinfo{person}{Nimrod
  Partush}, {and} \bibinfo{person}{Eran Yahav}.}
  \bibinfo{year}{2016}\natexlab{}.
\newblock \showarticletitle{Statistical similarity of binaries}.
\newblock \bibinfo{journal}{\emph{Acm sigplan notices}} \bibinfo{volume}{51},
  \bibinfo{number}{6} (\bibinfo{year}{2016}), \bibinfo{pages}{266--280}.
\newblock


\bibitem[David et~al\mbox{.}(2017)]%
        {Gitz}
\bibfield{author}{\bibinfo{person}{Yaniv David}, \bibinfo{person}{Nimrod
  Partush}, {and} \bibinfo{person}{Eran Yahav}.}
  \bibinfo{year}{2017}\natexlab{}.
\newblock \showarticletitle{Similarity of binaries through re-optimization}. In
  \bibinfo{booktitle}{\emph{Proceedings of the 38th ACM SIGPLAN conference on
  programming language design and implementation}}. \bibinfo{pages}{79--94}.
\newblock


\bibitem[David et~al\mbox{.}(2018)]%
        {Firmup}
\bibfield{author}{\bibinfo{person}{Yaniv David}, \bibinfo{person}{Nimrod
  Partush}, {and} \bibinfo{person}{Eran Yahav}.}
  \bibinfo{year}{2018}\natexlab{}.
\newblock \showarticletitle{FirmUp: Precise Static Detection of Common
  Vulnerabilities in Firmware}. In \bibinfo{booktitle}{\emph{Proceedings of the
  Twenty-Third International Conference on Architectural Support for
  Programming Languages and Operating Systems, {ASPLOS} 2018, Williamsburg, VA,
  USA, March 24-28, 2018}}, \bibfield{editor}{\bibinfo{person}{Xipeng Shen},
  \bibinfo{person}{James Tuck}, \bibinfo{person}{Ricardo Bianchini}, {and}
  \bibinfo{person}{Vivek Sarkar}} (Eds.). \bibinfo{publisher}{{ACM}},
  \bibinfo{pages}{392--404}.
\newblock
\urldef\tempurl%
\url{https://doi.org/10.1145/3173162.3177157}
\showDOI{\tempurl}


\bibitem[David and Yahav(2014)]%
        {Tracy}
\bibfield{author}{\bibinfo{person}{Yaniv David} {and} \bibinfo{person}{Eran
  Yahav}.} \bibinfo{year}{2014}\natexlab{}.
\newblock \showarticletitle{Tracelet-based code search in executables}.
\newblock \bibinfo{journal}{\emph{Acm Sigplan Notices}} \bibinfo{volume}{49},
  \bibinfo{number}{6} (\bibinfo{year}{2014}), \bibinfo{pages}{349--360}.
\newblock


\bibitem[Davidson and Holler(1992)]%
        {DBLP:journals/tse/DavidsonH92}
\bibfield{author}{\bibinfo{person}{Jack~W. Davidson} {and}
  \bibinfo{person}{Anne~M. Holler}.} \bibinfo{year}{1992}\natexlab{}.
\newblock \showarticletitle{Subprogram Inlining: {A} Study of its Effects on
  Program Execution Time}.
\newblock \bibinfo{journal}{\emph{{IEEE} Trans. Software Eng.}}
  \bibinfo{volume}{18}, \bibinfo{number}{2} (\bibinfo{year}{1992}),
  \bibinfo{pages}{89--102}.
\newblock
\urldef\tempurl%
\url{https://doi.org/10.1109/32.121752}
\showDOI{\tempurl}


\bibitem[Dean and Chambers(1994)]%
        {DBLP:conf/lfp/DeanC94}
\bibfield{author}{\bibinfo{person}{Jeffrey Dean} {and} \bibinfo{person}{Craig
  Chambers}.} \bibinfo{year}{1994}\natexlab{}.
\newblock \showarticletitle{Towards Better Inlining Decisions Using Inlining
  Trials}. In \bibinfo{booktitle}{\emph{Proceedings of the 1994 {ACM}
  Conference on {LISP} and Functional Programming, Orlando, Florida, USA, 27-29
  June 1994}}. \bibinfo{publisher}{{ACM}}, \bibinfo{pages}{273--282}.
\newblock
\urldef\tempurl%
\url{https://doi.org/10.1145/182409.182489}
\showDOI{\tempurl}


\bibitem[Ding et~al\mbox{.}(2016)]%
        {Kam1n0}
\bibfield{author}{\bibinfo{person}{Steven~HH Ding},
  \bibinfo{person}{Benjamin~CM Fung}, {and} \bibinfo{person}{Philippe
  Charland}.} \bibinfo{year}{2016}\natexlab{}.
\newblock \showarticletitle{Kam1n0: Mapreduce-based assembly clone search for
  reverse engineering}. In \bibinfo{booktitle}{\emph{Proceedings of the 22nd
  ACM SIGKDD international conference on knowledge discovery and data mining}}.
  \bibinfo{pages}{461--470}.
\newblock


\bibitem[Ding et~al\mbox{.}(2019)]%
        {asm2vec}
\bibfield{author}{\bibinfo{person}{Steven~HH Ding},
  \bibinfo{person}{Benjamin~CM Fung}, {and} \bibinfo{person}{Philippe
  Charland}.} \bibinfo{year}{2019}\natexlab{}.
\newblock \showarticletitle{Asm2vec: Boosting static representation robustness
  for binary clone search against code obfuscation and compiler optimization}.
  In \bibinfo{booktitle}{\emph{2019 IEEE Symposium on Security and Privacy
  (SP)}}. IEEE, \bibinfo{pages}{472--489}.
\newblock


\bibitem[Duan et~al\mbox{.}(2017)]%
        {OSSPolice}
\bibfield{author}{\bibinfo{person}{Ruian Duan}, \bibinfo{person}{Ashish
  Bijlani}, \bibinfo{person}{Meng Xu}, \bibinfo{person}{Taesoo Kim}, {and}
  \bibinfo{person}{Wenke Lee}.} \bibinfo{year}{2017}\natexlab{}.
\newblock \showarticletitle{Identifying Open-Source License Violation and 1-day
  Security Risk at Large Scale}. In \bibinfo{booktitle}{\emph{Proceedings of
  the 2017 {ACM} {SIGSAC} Conference on Computer and Communications Security,
  {CCS} 2017, Dallas, TX, USA, October 30 - November 03, 2017}}.
  \bibinfo{publisher}{{ACM}}, \bibinfo{pages}{2169--2185}.
\newblock
\urldef\tempurl%
\url{https://doi.org/10.1145/3133956.3134048}
\showDOI{\tempurl}


\bibitem[Duan et~al\mbox{.}(2020)]%
        {deepbindiff}
\bibfield{author}{\bibinfo{person}{Yue Duan}, \bibinfo{person}{Xuezixiang Li},
  \bibinfo{person}{Jinghan Wang}, {and} \bibinfo{person}{Heng Yin}.}
  \bibinfo{year}{2020}\natexlab{}.
\newblock \showarticletitle{Deepbindiff: Learning program-wide code
  representations for binary diffing}. In \bibinfo{booktitle}{\emph{Network and
  Distributed System Security Symposium}}.
\newblock


\bibitem[Durand and Strandh(2018)]%
        {durand2018partial}
\bibfield{author}{\bibinfo{person}{Ir{\`e}ne~A Durand} {and}
  \bibinfo{person}{Robert Strandh}.} \bibinfo{year}{2018}\natexlab{}.
\newblock \showarticletitle{Partial Inlining Using Local Graph Rewriting}. In
  \bibinfo{booktitle}{\emph{11th European Lisp Symposium}}.
\newblock


\bibitem[Egele et~al\mbox{.}(2014)]%
        {DBLP:conf/uss/EgeleWCB14}
\bibfield{author}{\bibinfo{person}{Manuel Egele}, \bibinfo{person}{Maverick
  Woo}, \bibinfo{person}{Peter Chapman}, {and} \bibinfo{person}{David
  Brumley}.} \bibinfo{year}{2014}\natexlab{}.
\newblock \showarticletitle{Blanket Execution: Dynamic Similarity Testing for
  Program Binaries and Components}. In \bibinfo{booktitle}{\emph{Proceedings of
  the 23rd {USENIX} Security Symposium, San Diego, CA, USA, August 20-22,
  2014}}, \bibfield{editor}{\bibinfo{person}{Kevin Fu} {and}
  \bibinfo{person}{Jaeyeon Jung}} (Eds.). \bibinfo{publisher}{{USENIX}
  Association}, \bibinfo{pages}{303--317}.
\newblock
\urldef\tempurl%
\url{https://www.usenix.org/conference/usenixsecurity14/technical-sessions/presentation/egele}
\showURL{%
\tempurl}


\bibitem[Feng et~al\mbox{.}(2019a)]%
        {Saner2019}
\bibfield{author}{\bibinfo{person}{Muyue Feng}, \bibinfo{person}{Weixuan Mao},
  \bibinfo{person}{Zimu Yuan}, \bibinfo{person}{Yang Xiao}, \bibinfo{person}{Gu
  Ban}, \bibinfo{person}{Wei Wang}, \bibinfo{person}{Shiyang Wang},
  \bibinfo{person}{Qian Tang}, \bibinfo{person}{Jiahuan Xu},
  \bibinfo{person}{He Su}, \bibinfo{person}{Binghong Liu}, {and}
  \bibinfo{person}{Wei Huo}.} \bibinfo{year}{2019}\natexlab{a}.
\newblock \showarticletitle{Open-Source License Violations of Binary Software
  at Large Scale}. In \bibinfo{booktitle}{\emph{26th {IEEE} International
  Conference on Software Analysis, Evolution and Reengineering, {SANER} 2019,
  Hangzhou, China, February 24-27, 2019}}. \bibinfo{publisher}{{IEEE}},
  \bibinfo{pages}{564--568}.
\newblock
\urldef\tempurl%
\url{https://doi.org/10.1109/SANER.2019.8667977}
\showDOI{\tempurl}


\bibitem[Feng et~al\mbox{.}(2019b)]%
        {B2SFinder}
\bibfield{author}{\bibinfo{person}{Muyue Feng}, \bibinfo{person}{Zimu Yuan},
  \bibinfo{person}{Feng Li}, \bibinfo{person}{Gu Ban}, \bibinfo{person}{Shiyang
  Wang}, \bibinfo{person}{Qian Tang}, \bibinfo{person}{He Su},
  \bibinfo{person}{Chendong Yu}, \bibinfo{person}{Jiahuan Xu},
  \bibinfo{person}{Aihua Piao}, \bibinfo{person}{Jingling Xue}, {and}
  \bibinfo{person}{Wei Huo}.} \bibinfo{year}{2019}\natexlab{b}.
\newblock \showarticletitle{B2SFinder: Detecting Open-Source Software Reuse in
  {COTS} Software}. In \bibinfo{booktitle}{\emph{34th {IEEE/ACM} International
  Conference on Automated Software Engineering, {ASE} 2019, San Diego, CA, USA,
  November 11-15, 2019}}. \bibinfo{publisher}{{IEEE}},
  \bibinfo{pages}{1038--1049}.
\newblock
\urldef\tempurl%
\url{https://doi.org/10.1109/ASE.2019.00100}
\showDOI{\tempurl}


\bibitem[Feng et~al\mbox{.}(2016)]%
        {Genius}
\bibfield{author}{\bibinfo{person}{Qian Feng}, \bibinfo{person}{Rundong Zhou},
  \bibinfo{person}{Chengcheng Xu}, \bibinfo{person}{Yao Cheng},
  \bibinfo{person}{Brian Testa}, {and} \bibinfo{person}{Heng Yin}.}
  \bibinfo{year}{2016}\natexlab{}.
\newblock \showarticletitle{Scalable graph-based bug search for firmware
  images}. In \bibinfo{booktitle}{\emph{Proceedings of the 2016 ACM SIGSAC
  Conference on Computer and Communications Security}}.
  \bibinfo{pages}{480--491}.
\newblock


\bibitem[Gao et~al\mbox{.}(2018)]%
        {vulseeker}
\bibfield{author}{\bibinfo{person}{Jian Gao}, \bibinfo{person}{Xin Yang},
  \bibinfo{person}{Ying Fu}, \bibinfo{person}{Yu Jiang}, {and}
  \bibinfo{person}{Jiaguang Sun}.} \bibinfo{year}{2018}\natexlab{}.
\newblock \showarticletitle{VulSeeker: a semantic learning based vulnerability
  seeker for cross-platform binary}. In \bibinfo{booktitle}{\emph{2018 33rd
  IEEE/ACM International Conference on Automated Software Engineering (ASE)}}.
  IEEE, \bibinfo{pages}{896--899}.
\newblock


\bibitem[Gharehyazie et~al\mbox{.}(2017)]%
        {CrossProjectReuse}
\bibfield{author}{\bibinfo{person}{Mohammad Gharehyazie},
  \bibinfo{person}{Baishakhi Ray}, {and} \bibinfo{person}{Vladimir Filkov}.}
  \bibinfo{year}{2017}\natexlab{}.
\newblock \showarticletitle{Some from here, some from there: Cross-project code
  reuse in github}. In \bibinfo{booktitle}{\emph{2017 IEEE/ACM 14th
  International Conference on Mining Software Repositories (MSR)}}. IEEE,
  \bibinfo{pages}{291--301}.
\newblock


\bibitem[Gkortzis et~al\mbox{.}(2021)]%
        {gkortzis2021software}
\bibfield{author}{\bibinfo{person}{Antonios Gkortzis}, \bibinfo{person}{Daniel
  Feitosa}, {and} \bibinfo{person}{Diomidis Spinellis}.}
  \bibinfo{year}{2021}\natexlab{}.
\newblock \showarticletitle{Software reuse cuts both ways: An empirical
  analysis of its relationship with security vulnerabilities}.
\newblock \bibinfo{journal}{\emph{Journal of Systems and Software}}
  \bibinfo{volume}{172} (\bibinfo{year}{2021}), \bibinfo{pages}{110653}.
\newblock


\bibitem[Hagberg et~al\mbox{.}(2008)]%
        {hagberg2008exploring}
\bibfield{author}{\bibinfo{person}{Aric Hagberg}, \bibinfo{person}{Pieter
  Swart}, {and} \bibinfo{person}{Daniel S~Chult}.}
  \bibinfo{year}{2008}\natexlab{}.
\newblock \bibinfo{booktitle}{\emph{Exploring network structure, dynamics, and
  function using NetworkX}}.
\newblock \bibinfo{type}{{T}echnical {R}eport}. \bibinfo{institution}{Los
  Alamos National Lab.(LANL), Los Alamos, NM (United States)}.
\newblock


\bibitem[Haq and Caballero(2021)]%
        {binary_similarity_survey}
\bibfield{author}{\bibinfo{person}{Irfan~Ul Haq} {and} \bibinfo{person}{Juan
  Caballero}.} \bibinfo{year}{2021}\natexlab{}.
\newblock \showarticletitle{A survey of binary code similarity}.
\newblock \bibinfo{journal}{\emph{ACM Computing Surveys (CSUR)}}
  \bibinfo{volume}{54}, \bibinfo{number}{3} (\bibinfo{year}{2021}),
  \bibinfo{pages}{1--38}.
\newblock


\bibitem[Hemel et~al\mbox{.}(2011)]%
        {BAT}
\bibfield{author}{\bibinfo{person}{Armijn Hemel}, \bibinfo{person}{Karl~Trygve
  Kalleberg}, \bibinfo{person}{Rob Vermaas}, {and} \bibinfo{person}{Eelco
  Dolstra}.} \bibinfo{year}{2011}\natexlab{}.
\newblock \showarticletitle{Finding software license violations through binary
  code clone detection}. In \bibinfo{booktitle}{\emph{Proceedings of the 8th
  International Working Conference on Mining Software Repositories, {MSR} 2011
  (Co-located with ICSE), Waikiki, Honolulu, HI, USA, May 21-28, 2011,
  Proceedings}}. \bibinfo{publisher}{{ACM}}, \bibinfo{pages}{63--72}.
\newblock
\urldef\tempurl%
\url{https://doi.org/10.1145/1985441.1985453}
\showDOI{\tempurl}


\bibitem[Hu et~al\mbox{.}(2016)]%
        {DBLP:conf/wcre/HuZLG16}
\bibfield{author}{\bibinfo{person}{Yikun Hu}, \bibinfo{person}{Yuanyuan Zhang},
  \bibinfo{person}{Juanru Li}, {and} \bibinfo{person}{Dawu Gu}.}
  \bibinfo{year}{2016}\natexlab{}.
\newblock \showarticletitle{Cross-Architecture Binary Semantics Understanding
  via Similar Code Comparison}. In \bibinfo{booktitle}{\emph{{IEEE} 23rd
  International Conference on Software Analysis, Evolution, and Reengineering,
  {SANER} 2016, Suita, Osaka, Japan, March 14-18, 2016 - Volume 1}}.
  \bibinfo{publisher}{{IEEE} Computer Society}, \bibinfo{pages}{57--67}.
\newblock
\urldef\tempurl%
\url{https://doi.org/10.1109/SANER.2016.50}
\showDOI{\tempurl}


\bibitem[Hubicka(2004)]%
        {hubicka2004gcc}
\bibfield{author}{\bibinfo{person}{Jan Hubicka}.}
  \bibinfo{year}{2004}\natexlab{}.
\newblock \bibinfo{title}{The GCC call graph module: a framework for
  inter-procedural optimization}.
\newblock
\newblock


\bibitem[Hwu and Chang(1989)]%
        {DBLP:conf/pldi/HwuC89}
\bibfield{author}{\bibinfo{person}{Wen{-}mei~W. Hwu} {and}
  \bibinfo{person}{Pohua~P. Chang}.} \bibinfo{year}{1989}\natexlab{}.
\newblock \showarticletitle{Inline Function Expansion for Compiling {C}
  Programs}. In \bibinfo{booktitle}{\emph{Proceedings of the {ACM} SIGPLAN'89
  Conference on Programming Language Design and Implementation (PLDI),
  Portland, Oregon, USA, June 21-23, 1989}}. \bibinfo{publisher}{{ACM}},
  \bibinfo{pages}{246--257}.
\newblock
\urldef\tempurl%
\url{https://doi.org/10.1145/73141.74840}
\showDOI{\tempurl}


\bibitem[Jang et~al\mbox{.}(2013)]%
        {DBLP:conf/uss/JangWB13}
\bibfield{author}{\bibinfo{person}{Jiyong Jang}, \bibinfo{person}{Maverick
  Woo}, {and} \bibinfo{person}{David Brumley}.}
  \bibinfo{year}{2013}\natexlab{}.
\newblock \showarticletitle{Towards Automatic Software Lineage Inference}. In
  \bibinfo{booktitle}{\emph{Proceedings of the 22th {USENIX} Security
  Symposium, Washington, DC, USA, August 14-16, 2013}},
  \bibfield{editor}{\bibinfo{person}{Samuel~T. King}} (Ed.).
  \bibinfo{publisher}{{USENIX} Association}, \bibinfo{pages}{81--96}.
\newblock
\urldef\tempurl%
\url{https://www.usenix.org/conference/usenixsecurity13/technical-sessions/papers/jang}
\showURL{%
\tempurl}


\bibitem[Jia et~al\mbox{.}({[n.\,d.]})]%
        {jia20221}
\bibfield{author}{\bibinfo{person}{Ang Jia}, \bibinfo{person}{Ming Fan},
  \bibinfo{person}{Wuxia Jin}, \bibinfo{person}{Xi Xu},
  \bibinfo{person}{Zhaohui Zhou}, \bibinfo{person}{Qiyi Tang},
  \bibinfo{person}{Sen Nie}, \bibinfo{person}{Shi Wu}, {and}
  \bibinfo{person}{Ting Liu}.} \bibinfo{year}{[n.\,d.]}\natexlab{}.
\newblock \showarticletitle{1-to-1 or 1-to-n? Investigating the effect of
  function inlining on binary similarity analysis}.
\newblock \bibinfo{journal}{\emph{ACM Transactions on Software Engineering and
  Methodology}} (\bibinfo{year}{[n.\,d.]}).
\newblock


\bibitem[Jia et~al\mbox{.}(2023)]%
        {Tosem}
\bibfield{author}{\bibinfo{person}{Ang Jia}, \bibinfo{person}{Ming Fan},
  \bibinfo{person}{Wuxia Jin}, \bibinfo{person}{Xi Xu},
  \bibinfo{person}{Zhaohui Zhou}, \bibinfo{person}{Qiyi Tang},
  \bibinfo{person}{Sen Nie}, \bibinfo{person}{Shi Wu}, {and}
  \bibinfo{person}{Ting Liu}.} \bibinfo{year}{2023}\natexlab{}.
\newblock \showarticletitle{1-to-1 or 1-to-n? Investigating the Effect of
  Function Inlining on Binary Similarity Analysis}.
\newblock \bibinfo{journal}{\emph{ACM Trans. Softw. Eng. Methodol.}}
  \bibinfo{volume}{32}, \bibinfo{number}{4}, Article \bibinfo{articleno}{87}
  (\bibinfo{date}{may} \bibinfo{year}{2023}), \bibinfo{numpages}{26}~pages.
\newblock
\showISSN{1049-331X}
\urldef\tempurl%
\url{https://doi.org/10.1145/3561385}
\showDOI{\tempurl}


\bibitem[Jia et~al\mbox{.}(2022)]%
        {jia2022comparing}
\bibfield{author}{\bibinfo{person}{Ang Jia}, \bibinfo{person}{Ming Fan},
  \bibinfo{person}{Xi Xu}, \bibinfo{person}{Wuxia Jin}, \bibinfo{person}{Haijun
  Wang}, \bibinfo{person}{Qiyi Tang}, \bibinfo{person}{Sen Nie},
  \bibinfo{person}{Shi Wu}, {and} \bibinfo{person}{Ting Liu}.}
  \bibinfo{year}{2022}\natexlab{}.
\newblock \showarticletitle{Comparing One with Many--Solving Binary2source
  Function Matching Under Function Inlining}.
\newblock \bibinfo{journal}{\emph{arXiv preprint arXiv:2210.15159}}
  (\bibinfo{year}{2022}).
\newblock


\bibitem[Jiang et~al\mbox{.}(2020)]%
        {PDiff}
\bibfield{author}{\bibinfo{person}{Zheyue Jiang}, \bibinfo{person}{Yuan Zhang},
  \bibinfo{person}{Jun Xu}, \bibinfo{person}{Qi Wen}, \bibinfo{person}{Zhenghe
  Wang}, \bibinfo{person}{Xiaohan Zhang}, \bibinfo{person}{Xinyu Xing},
  \bibinfo{person}{Min Yang}, {and} \bibinfo{person}{Zhemin Yang}.}
  \bibinfo{year}{2020}\natexlab{}.
\newblock \showarticletitle{PDiff: Semantic-based Patch Presence Testing for
  Downstream Kernels}. In \bibinfo{booktitle}{\emph{{CCS} '20: 2020 {ACM}
  {SIGSAC} Conference on Computer and Communications Security, Virtual Event,
  USA, November 9-13, 2020}}, \bibfield{editor}{\bibinfo{person}{Jay Ligatti},
  \bibinfo{person}{Xinming Ou}, \bibinfo{person}{Jonathan Katz}, {and}
  \bibinfo{person}{Giovanni Vigna}} (Eds.). \bibinfo{publisher}{{ACM}},
  \bibinfo{pages}{1149--1163}.
\newblock
\urldef\tempurl%
\url{https://doi.org/10.1145/3372297.3417240}
\showDOI{\tempurl}


\bibitem[Karg{\'{e}}n and Shahmehri(2017)]%
        {DBLP:conf/kbse/KargenS17}
\bibfield{author}{\bibinfo{person}{Ulf Karg{\'{e}}n} {and}
  \bibinfo{person}{Nahid Shahmehri}.} \bibinfo{year}{2017}\natexlab{}.
\newblock \showarticletitle{Towards robust instruction-level trace alignment of
  binary code}. In \bibinfo{booktitle}{\emph{Proceedings of the 32nd {IEEE/ACM}
  International Conference on Automated Software Engineering, {ASE} 2017,
  Urbana, IL, USA, October 30 - November 03, 2017}},
  \bibfield{editor}{\bibinfo{person}{Grigore Rosu},
  \bibinfo{person}{Massimiliano~Di Penta}, {and} \bibinfo{person}{Tien~N.
  Nguyen}} (Eds.). \bibinfo{publisher}{{IEEE} Computer Society},
  \bibinfo{pages}{342--352}.
\newblock
\urldef\tempurl%
\url{https://doi.org/10.1109/ASE.2017.8115647}
\showDOI{\tempurl}


\bibitem[Kim et~al\mbox{.}(2014)]%
        {JISIS14}
\bibfield{author}{\bibinfo{person}{Dongjin Kim}, \bibinfo{person}{Seong{-}je
  Cho}, \bibinfo{person}{Sangchul Han}, \bibinfo{person}{Minkyu Park}, {and}
  \bibinfo{person}{Ilsun You}.} \bibinfo{year}{2014}\natexlab{}.
\newblock \showarticletitle{Open Source Software Detection using Function-level
  Static Software Birthmark}.
\newblock \bibinfo{journal}{\emph{J. Internet Serv. Inf. Secur.}}
  \bibinfo{volume}{4}, \bibinfo{number}{4} (\bibinfo{year}{2014}),
  \bibinfo{pages}{25--37}.
\newblock
\urldef\tempurl%
\url{https://doi.org/10.22667/JISIS.2014.11.31.025}
\showDOI{\tempurl}


\bibitem[Kim et~al\mbox{.}(2020)]%
        {Binkit}
\bibfield{author}{\bibinfo{person}{Dongkwan Kim}, \bibinfo{person}{Eunsoo Kim},
  \bibinfo{person}{Sang~Kil Cha}, \bibinfo{person}{Sooel Son}, {and}
  \bibinfo{person}{Yongdae Kim}.} \bibinfo{year}{2020}\natexlab{}.
\newblock \showarticletitle{Revisiting Binary Code Similarity Analysis using
  Interpretable Feature Engineering and Lessons Learned}.
\newblock  (\bibinfo{year}{2020}).
\newblock
\showeprint[arxiv]{2011.10749}~[cs.SE]


\bibitem[Kim et~al\mbox{.}(2019)]%
        {DBLP:journals/tjs/KimLKI19}
\bibfield{author}{\bibinfo{person}{TaeGuen Kim}, \bibinfo{person}{Yeo~Reum
  Lee}, \bibinfo{person}{BooJoong Kang}, {and} \bibinfo{person}{Eul~Gyu Im}.}
  \bibinfo{year}{2019}\natexlab{}.
\newblock \showarticletitle{Binary executable file similarity calculation using
  function matching}.
\newblock \bibinfo{journal}{\emph{J. Supercomput.}} \bibinfo{volume}{75},
  \bibinfo{number}{2} (\bibinfo{year}{2019}), \bibinfo{pages}{607--622}.
\newblock
\urldef\tempurl%
\url{https://doi.org/10.1007/s11227-016-1941-2}
\showDOI{\tempurl}


\bibitem[Kula et~al\mbox{.}(2018)]%
        {kula2018developers}
\bibfield{author}{\bibinfo{person}{Raula~Gaikovina Kula},
  \bibinfo{person}{Daniel~M German}, \bibinfo{person}{Ali Ouni},
  \bibinfo{person}{Takashi Ishio}, {and} \bibinfo{person}{Katsuro Inoue}.}
  \bibinfo{year}{2018}\natexlab{}.
\newblock \showarticletitle{Do developers update their library dependencies?}
\newblock \bibinfo{journal}{\emph{Empirical Software Engineering}}
  \bibinfo{volume}{23}, \bibinfo{number}{1} (\bibinfo{year}{2018}),
  \bibinfo{pages}{384--417}.
\newblock


\bibitem[Li et~al\mbox{.}(2019)]%
        {GMN_matching}
\bibfield{author}{\bibinfo{person}{Yujia Li}, \bibinfo{person}{Chenjie Gu},
  \bibinfo{person}{Thomas Dullien}, \bibinfo{person}{Oriol Vinyals}, {and}
  \bibinfo{person}{Pushmeet Kohli}.} \bibinfo{year}{2019}\natexlab{}.
\newblock \showarticletitle{Graph Matching Networks for Learning the Similarity
  of Graph Structured Objects}. In \bibinfo{booktitle}{\emph{Proceedings of the
  36th International Conference on Machine Learning, {ICML} 2019, 9-15 June
  2019, Long Beach, California, {USA}}} \emph{(\bibinfo{series}{Proceedings of
  Machine Learning Research}, Vol.~\bibinfo{volume}{97})},
  \bibfield{editor}{\bibinfo{person}{Kamalika Chaudhuri} {and}
  \bibinfo{person}{Ruslan Salakhutdinov}} (Eds.). \bibinfo{publisher}{{PMLR}},
  \bibinfo{pages}{3835--3845}.
\newblock
\urldef\tempurl%
\url{http://proceedings.mlr.press/v97/li19d.html}
\showURL{%
\tempurl}


\bibitem[Lin et~al\mbox{.}(2017)]%
        {lin2017structured}
\bibfield{author}{\bibinfo{person}{Zhouhan Lin}, \bibinfo{person}{Minwei Feng},
  \bibinfo{person}{Cicero Nogueira~dos Santos}, \bibinfo{person}{Mo Yu},
  \bibinfo{person}{Bing Xiang}, \bibinfo{person}{Bowen Zhou}, {and}
  \bibinfo{person}{Yoshua Bengio}.} \bibinfo{year}{2017}\natexlab{}.
\newblock \showarticletitle{A structured self-attentive sentence embedding}.
\newblock \bibinfo{journal}{\emph{arXiv preprint arXiv:1703.03130}}
  (\bibinfo{year}{2017}).
\newblock


\bibitem[Lindorfer et~al\mbox{.}(2012)]%
        {DBLP:conf/acsac/LindorferFMCZ12}
\bibfield{author}{\bibinfo{person}{Martina Lindorfer},
  \bibinfo{person}{Alessandro~Di Federico}, \bibinfo{person}{Federico Maggi},
  \bibinfo{person}{Paolo~Milani Comparetti}, {and} \bibinfo{person}{Stefano
  Zanero}.} \bibinfo{year}{2012}\natexlab{}.
\newblock \showarticletitle{Lines of malicious code: insights into the
  malicious software industry}. In \bibinfo{booktitle}{\emph{28th Annual
  Computer Security Applications Conference, {ACSAC} 2012, Orlando, FL, USA,
  3-7 December 2012}}, \bibfield{editor}{\bibinfo{person}{Robert~H'obbes'
  Zakon}} (Ed.). \bibinfo{publisher}{{ACM}}, \bibinfo{pages}{349--358}.
\newblock
\urldef\tempurl%
\url{https://doi.org/10.1145/2420950.2421001}
\showDOI{\tempurl}


\bibitem[Liu et~al\mbox{.}(2018)]%
        {alphadiff}
\bibfield{author}{\bibinfo{person}{Bingchang Liu}, \bibinfo{person}{Wei Huo},
  \bibinfo{person}{Chao Zhang}, \bibinfo{person}{Wenchao Li},
  \bibinfo{person}{Feng Li}, \bibinfo{person}{Aihua Piao}, {and}
  \bibinfo{person}{Wei Zou}.} \bibinfo{year}{2018}\natexlab{}.
\newblock \showarticletitle{$\alpha$diff: cross-version binary code similarity
  detection with dnn}. In \bibinfo{booktitle}{\emph{Proceedings of the 33rd
  ACM/IEEE International Conference on Automated Software Engineering}}.
  \bibinfo{pages}{667--678}.
\newblock


\bibitem[Liu et~al\mbox{.}(2022)]%
        {PG-VulNet}
\bibfield{author}{\bibinfo{person}{Xin Liu}, \bibinfo{person}{Yixiong Wu},
  \bibinfo{person}{Qingchen Yu}, \bibinfo{person}{Shangru Song},
  \bibinfo{person}{Yue Liu}, \bibinfo{person}{Qingguo Zhou}, {and}
  \bibinfo{person}{Jianwei Zhuge}.} \bibinfo{year}{2022}\natexlab{}.
\newblock \showarticletitle{PG-VulNet: Detect Supply Chain Vulnerabilities in
  IoT Devices using Pseudo-code and Graphs}. In
  \bibinfo{booktitle}{\emph{{ESEM} '22: {ACM} / {IEEE} International Symposium
  on Empirical Software Engineering and Measurement, Helsinki Finland,
  September 19 - 23, 2022}}, \bibfield{editor}{\bibinfo{person}{Fernanda
  Madeiral}, \bibinfo{person}{Casper Lassenius}, \bibinfo{person}{Tayana
  Conte}, {and} \bibinfo{person}{Tomi M{\"{a}}nnist{\"{o}}}} (Eds.).
  \bibinfo{publisher}{{ACM}}, \bibinfo{pages}{205--215}.
\newblock
\urldef\tempurl%
\url{https://doi.org/10.1145/3544902.3546240}
\showDOI{\tempurl}


\bibitem[Marcelli et~al\mbox{.}(2022)]%
        {b2b_evaluation}
\bibfield{author}{\bibinfo{person}{Andrea Marcelli}, \bibinfo{person}{Mariano
  Graziano}, \bibinfo{person}{Xabier Ugarte-Pedrero}, \bibinfo{person}{Yanick
  Fratantonio}, \bibinfo{person}{Mohamad Mansouri}, {and}
  \bibinfo{person}{Davide Balzarotti}.} \bibinfo{year}{2022}\natexlab{}.
\newblock \showarticletitle{How machine learning is solving the binary function
  similarity problem}. In \bibinfo{booktitle}{\emph{31st USENIX Security
  Symposium (USENIX Security 22)}}. \bibinfo{pages}{2099--2116}.
\newblock


\bibitem[Massarelli et~al\mbox{.}(2019)]%
        {safe}
\bibfield{author}{\bibinfo{person}{Luca Massarelli},
  \bibinfo{person}{Giuseppe~Antonio Di~Luna}, \bibinfo{person}{Fabio Petroni},
  \bibinfo{person}{Roberto Baldoni}, {and} \bibinfo{person}{Leonardo
  Querzoni}.} \bibinfo{year}{2019}\natexlab{}.
\newblock \showarticletitle{Safe: Self-attentive function embeddings for binary
  similarity}. In \bibinfo{booktitle}{\emph{International Conference on
  Detection of Intrusions and Malware, and Vulnerability Assessment}}.
  Springer, \bibinfo{pages}{309--329}.
\newblock


\bibitem[McCallum et~al\mbox{.}(1998)]%
        {bag_of_words}
\bibfield{author}{\bibinfo{person}{Andrew McCallum}, \bibinfo{person}{Kamal
  Nigam}, {et~al\mbox{.}}} \bibinfo{year}{1998}\natexlab{}.
\newblock \showarticletitle{A comparison of event models for naive bayes text
  classification}. In \bibinfo{booktitle}{\emph{AAAI-98 workshop on learning
  for text categorization}}, Vol.~\bibinfo{volume}{752}. Madison, WI,
  \bibinfo{pages}{41--48}.
\newblock


\bibitem[Ming et~al\mbox{.}(2012)]%
        {DBLP:conf/icisc/MingPG12}
\bibfield{author}{\bibinfo{person}{Jiang Ming}, \bibinfo{person}{Meng Pan},
  {and} \bibinfo{person}{Debin Gao}.} \bibinfo{year}{2012}\natexlab{}.
\newblock \showarticletitle{iBinHunt: Binary Hunting with Inter-procedural
  Control Flow}. In \bibinfo{booktitle}{\emph{Information Security and
  Cryptology - {ICISC} 2012 - 15th International Conference, Seoul, Korea,
  November 28-30, 2012, Revised Selected Papers}}
  \emph{(\bibinfo{series}{Lecture Notes in Computer Science},
  Vol.~\bibinfo{volume}{7839})}, \bibfield{editor}{\bibinfo{person}{Taekyoung
  Kwon}, \bibinfo{person}{Mun{-}Kyu Lee}, {and} \bibinfo{person}{Daesung Kwon}}
  (Eds.). \bibinfo{publisher}{Springer}, \bibinfo{pages}{92--109}.
\newblock
\urldef\tempurl%
\url{https://doi.org/10.1007/978-3-642-37682-5\_8}
\showDOI{\tempurl}


\bibitem[Ming et~al\mbox{.}(2017)]%
        {DBLP:conf/uss/MingXJW17}
\bibfield{author}{\bibinfo{person}{Jiang Ming}, \bibinfo{person}{Dongpeng Xu},
  \bibinfo{person}{Yufei Jiang}, {and} \bibinfo{person}{Dinghao Wu}.}
  \bibinfo{year}{2017}\natexlab{}.
\newblock \showarticletitle{BinSim: Trace-based Semantic Binary Diffing via
  System Call Sliced Segment Equivalence Checking}. In
  \bibinfo{booktitle}{\emph{26th {USENIX} Security Symposium, {USENIX} Security
  2017, Vancouver, BC, Canada, August 16-18, 2017}},
  \bibfield{editor}{\bibinfo{person}{Engin Kirda} {and} \bibinfo{person}{Thomas
  Ristenpart}} (Eds.). \bibinfo{publisher}{{USENIX} Association},
  \bibinfo{pages}{253--270}.
\newblock
\urldef\tempurl%
\url{https://www.usenix.org/conference/usenixsecurity17/technical-sessions/presentation/ming}
\showURL{%
\tempurl}


\bibitem[Ming et~al\mbox{.}(2015)]%
        {DBLP:conf/sec/MingXW15}
\bibfield{author}{\bibinfo{person}{Jiang Ming}, \bibinfo{person}{Dongpeng Xu},
  {and} \bibinfo{person}{Dinghao Wu}.} \bibinfo{year}{2015}\natexlab{}.
\newblock \showarticletitle{Memoized Semantics-Based Binary Diffing with
  Application to Malware Lineage Inference}. In \bibinfo{booktitle}{\emph{{ICT}
  Systems Security and Privacy Protection - 30th {IFIP} {TC} 11 International
  Conference, {SEC} 2015, Hamburg, Germany, May 26-28, 2015, Proceedings}}
  \emph{(\bibinfo{series}{{IFIP} Advances in Information and Communication
  Technology}, Vol.~\bibinfo{volume}{455})},
  \bibfield{editor}{\bibinfo{person}{Hannes Federrath} {and}
  \bibinfo{person}{Dieter Gollmann}} (Eds.). \bibinfo{publisher}{Springer},
  \bibinfo{pages}{416--430}.
\newblock
\urldef\tempurl%
\url{https://doi.org/10.1007/978-3-319-18467-8\_28}
\showDOI{\tempurl}


\bibitem[Miyani et~al\mbox{.}(2017)]%
        {BinPro}
\bibfield{author}{\bibinfo{person}{Dhaval Miyani}, \bibinfo{person}{Zhen
  Huang}, {and} \bibinfo{person}{David Lie}.} \bibinfo{year}{2017}\natexlab{}.
\newblock \showarticletitle{BinPro: {A} Tool for Binary Source Code
  Provenance}.
\newblock \bibinfo{journal}{\emph{CoRR}}  \bibinfo{volume}{abs/1711.00830}
  (\bibinfo{year}{2017}).
\newblock
\showeprint[arxiv]{1711.00830}
\urldef\tempurl%
\url{http://arxiv.org/abs/1711.00830}
\showURL{%
\tempurl}


\bibitem[Pang et~al\mbox{.}(2021)]%
        {pang2021sok}
\bibfield{author}{\bibinfo{person}{Chengbin Pang}, \bibinfo{person}{Ruotong
  Yu}, \bibinfo{person}{Yaohui Chen}, \bibinfo{person}{Eric Koskinen},
  \bibinfo{person}{Georgios Portokalidis}, \bibinfo{person}{Bing Mao}, {and}
  \bibinfo{person}{Jun Xu}.} \bibinfo{year}{2021}\natexlab{}.
\newblock \showarticletitle{Sok: All you ever wanted to know about x86/x64
  binary disassembly but were afraid to ask}. In \bibinfo{booktitle}{\emph{2021
  IEEE Symposium on Security and Privacy (SP)}}. IEEE,
  \bibinfo{pages}{833--851}.
\newblock


\bibitem[Pei et~al\mbox{.}(2020)]%
        {trex}
\bibfield{author}{\bibinfo{person}{Kexin Pei}, \bibinfo{person}{Zhou Xuan},
  \bibinfo{person}{Junfeng Yang}, \bibinfo{person}{Suman Jana}, {and}
  \bibinfo{person}{Baishakhi Ray}.} \bibinfo{year}{2020}\natexlab{}.
\newblock \showarticletitle{Trex: Learning execution semantics from
  micro-traces for binary similarity}.
\newblock \bibinfo{journal}{\emph{arXiv preprint arXiv:2012.08680}}
  (\bibinfo{year}{2020}).
\newblock


\bibitem[Rahimian et~al\mbox{.}(2012)]%
        {RESource}
\bibfield{author}{\bibinfo{person}{Ashkan Rahimian}, \bibinfo{person}{Philippe
  Charland}, \bibinfo{person}{Stere Preda}, {and} \bibinfo{person}{Mourad
  Debbabi}.} \bibinfo{year}{2012}\natexlab{}.
\newblock \showarticletitle{RESource: {A} Framework for Online Matching of
  Assembly with Open Source Code}. In \bibinfo{booktitle}{\emph{Foundations and
  Practice of Security - 5th International Symposium, {FPS} 2012, Montreal, QC,
  Canada, October 25-26, 2012, Revised Selected Papers}}
  \emph{(\bibinfo{series}{Lecture Notes in Computer Science},
  Vol.~\bibinfo{volume}{7743})}. \bibinfo{publisher}{Springer},
  \bibinfo{pages}{211--226}.
\newblock
\urldef\tempurl%
\url{https://doi.org/10.1007/978-3-642-37119-6\_14}
\showDOI{\tempurl}


\bibitem[Redmond et~al\mbox{.}(2018)]%
        {RLZ2019}
\bibfield{author}{\bibinfo{person}{Kimberly Redmond}, \bibinfo{person}{Lannan
  Luo}, {and} \bibinfo{person}{Qiang Zeng}.} \bibinfo{year}{2018}\natexlab{}.
\newblock \showarticletitle{A cross-architecture instruction embedding model
  for natural language processing-inspired binary code analysis}.
\newblock \bibinfo{journal}{\emph{arXiv preprint arXiv:1812.09652}}
  (\bibinfo{year}{2018}).
\newblock


\bibitem[Ren et~al\mbox{.}(2021)]%
        {10.1145/3453483.3454035}
\bibfield{author}{\bibinfo{person}{Xiaolei Ren}, \bibinfo{person}{Michael Ho},
  \bibinfo{person}{Jiang Ming}, \bibinfo{person}{Yu Lei}, {and}
  \bibinfo{person}{Li Li}.} \bibinfo{year}{2021}\natexlab{}.
\newblock \bibinfo{booktitle}{\emph{Unleashing the Hidden Power of Compiler
  Optimization on Binary Code Difference: An Empirical Study}}.
\newblock \bibinfo{publisher}{Association for Computing Machinery},
  \bibinfo{address}{New York, NY, USA}, \bibinfo{pages}{142–157}.
\newblock
\showISBNx{9781450383912}
\urldef\tempurl%
\url{https://doi.org/10.1145/3453483.3454035}
\showURL{%
\tempurl}


\bibitem[Sun et~al\mbox{.}(2021)]%
        {Osprey}
\bibfield{author}{\bibinfo{person}{Peiyuan Sun}, \bibinfo{person}{Qiben Yan},
  \bibinfo{person}{Haoyi Zhou}, {and} \bibinfo{person}{Jianxin Li}.}
  \bibinfo{year}{2021}\natexlab{}.
\newblock \showarticletitle{Osprey: {A} fast and accurate patch presence test
  framework for binaries}.
\newblock \bibinfo{journal}{\emph{Comput. Commun.}}  \bibinfo{volume}{173}
  (\bibinfo{year}{2021}), \bibinfo{pages}{95--106}.
\newblock
\urldef\tempurl%
\url{https://doi.org/10.1016/j.comcom.2021.03.011}
\showDOI{\tempurl}


\bibitem[Wang and Wu(2017)]%
        {DBLP:conf/kbse/WangW17}
\bibfield{author}{\bibinfo{person}{Shuai Wang} {and} \bibinfo{person}{Dinghao
  Wu}.} \bibinfo{year}{2017}\natexlab{}.
\newblock \showarticletitle{In-memory fuzzing for binary code similarity
  analysis}. In \bibinfo{booktitle}{\emph{Proceedings of the 32nd {IEEE/ACM}
  International Conference on Automated Software Engineering, {ASE} 2017,
  Urbana, IL, USA, October 30 - November 03, 2017}},
  \bibfield{editor}{\bibinfo{person}{Grigore Rosu},
  \bibinfo{person}{Massimiliano~Di Penta}, {and} \bibinfo{person}{Tien~N.
  Nguyen}} (Eds.). \bibinfo{publisher}{{IEEE} Computer Society},
  \bibinfo{pages}{319--330}.
\newblock
\urldef\tempurl%
\url{https://doi.org/10.1109/ASE.2017.8115645}
\showDOI{\tempurl}


\bibitem[Xu et~al\mbox{.}(2017)]%
        {Gemini}
\bibfield{author}{\bibinfo{person}{Xiaojun Xu}, \bibinfo{person}{Chang Liu},
  \bibinfo{person}{Qian Feng}, \bibinfo{person}{Heng Yin}, \bibinfo{person}{Le
  Song}, {and} \bibinfo{person}{Dawn Song}.} \bibinfo{year}{2017}\natexlab{}.
\newblock \showarticletitle{Neural network-based graph embedding for
  cross-platform binary code similarity detection}. In
  \bibinfo{booktitle}{\emph{Proceedings of the 2017 ACM SIGSAC Conference on
  Computer and Communications Security}}. \bibinfo{pages}{363--376}.
\newblock


\bibitem[Xu et~al\mbox{.}(2021)]%
        {ISRD}
\bibfield{author}{\bibinfo{person}{Xi Xu}, \bibinfo{person}{Qinghua Zheng},
  \bibinfo{person}{Zheng Yan}, \bibinfo{person}{Ming Fan}, \bibinfo{person}{Ang
  Jia}, {and} \bibinfo{person}{Ting Liu}.} \bibinfo{year}{2021}\natexlab{}.
\newblock \showarticletitle{Interpretation-enabled Software Reuse Detection
  Based on a Multi-Level Birthmark Model}. In \bibinfo{booktitle}{\emph{43rd
  {IEEE/ACM} International Conference on Software Engineering, {ICSE} 2021,
  Madrid, Spain, 22-30 May 2021}}. \bibinfo{publisher}{{IEEE}},
  \bibinfo{pages}{873--884}.
\newblock
\urldef\tempurl%
\url{https://doi.org/10.1109/ICSE43902.2021.00084}
\showDOI{\tempurl}


\bibitem[Xu et~al\mbox{.}(2020)]%
        {Binxray}
\bibfield{author}{\bibinfo{person}{Yifei Xu}, \bibinfo{person}{Zhengzi Xu},
  \bibinfo{person}{Bihuan Chen}, \bibinfo{person}{Fu Song},
  \bibinfo{person}{Yang Liu}, {and} \bibinfo{person}{Ting Liu}.}
  \bibinfo{year}{2020}\natexlab{}.
\newblock \showarticletitle{Patch based vulnerability matching for binary
  programs}. In \bibinfo{booktitle}{\emph{Proceedings of the 29th ACM SIGSOFT
  International Symposium on Software Testing and Analysis}}.
  \bibinfo{pages}{376--387}.
\newblock


\bibitem[Xue et~al\mbox{.}(2018)]%
        {bingo-E}
\bibfield{author}{\bibinfo{person}{Yinxing Xue}, \bibinfo{person}{Zhengzi Xu},
  \bibinfo{person}{Mahinthan Chandramohan}, {and} \bibinfo{person}{Yang Liu}.}
  \bibinfo{year}{2018}\natexlab{}.
\newblock \showarticletitle{Accurate and scalable cross-architecture cross-os
  binary code search with emulation}.
\newblock \bibinfo{journal}{\emph{IEEE Transactions on Software Engineering}}
  \bibinfo{volume}{45}, \bibinfo{number}{11} (\bibinfo{year}{2018}),
  \bibinfo{pages}{1125--1149}.
\newblock


\bibitem[Yu et~al\mbox{.}(2020a)]%
        {yu2020order}
\bibfield{author}{\bibinfo{person}{Zeping Yu}, \bibinfo{person}{Rui Cao},
  \bibinfo{person}{Qiyi Tang}, \bibinfo{person}{Sen Nie},
  \bibinfo{person}{Junzhou Huang}, {and} \bibinfo{person}{Shi Wu}.}
  \bibinfo{year}{2020}\natexlab{a}.
\newblock \showarticletitle{Order matters: Semantic-aware neural networks for
  binary code similarity detection}. In \bibinfo{booktitle}{\emph{Proceedings
  of the AAAI conference on artificial intelligence}},
  Vol.~\bibinfo{volume}{34}. \bibinfo{pages}{1145--1152}.
\newblock


\bibitem[Yu et~al\mbox{.}(2020b)]%
        {CodeCMR}
\bibfield{author}{\bibinfo{person}{Zeping Yu}, \bibinfo{person}{Wenxin Zheng},
  \bibinfo{person}{Jiaqi Wang}, \bibinfo{person}{Qiyi Tang},
  \bibinfo{person}{Sen Nie}, {and} \bibinfo{person}{Shi Wu}.}
  \bibinfo{year}{2020}\natexlab{b}.
\newblock \showarticletitle{CodeCMR: Cross-Modal Retrieval For Function-Level
  Binary Source Code Matching}. In \bibinfo{booktitle}{\emph{Advances in Neural
  Information Processing Systems 33: Annual Conference on Neural Information
  Processing Systems 2020, NeurIPS 2020, December 6-12, 2020, virtual}}.
\newblock


\bibitem[Zhang and Qian(2018)]%
        {FIBER}
\bibfield{author}{\bibinfo{person}{Hang Zhang} {and} \bibinfo{person}{Zhiyun
  Qian}.} \bibinfo{year}{2018}\natexlab{}.
\newblock \showarticletitle{Precise and Accurate Patch Presence Test for
  Binaries}. In \bibinfo{booktitle}{\emph{27th {USENIX} Security Symposium,
  {USENIX} Security 2018, Baltimore, MD, USA, August 15-17, 2018}}.
  \bibinfo{publisher}{{USENIX} Association}, \bibinfo{pages}{887--902}.
\newblock


\bibitem[Zhao and Amaral(2003)]%
        {DBLP:conf/lcpc/ZhaoA03}
\bibfield{author}{\bibinfo{person}{Peng Zhao} {and}
  \bibinfo{person}{Jos{\'{e}}~Nelson Amaral}.} \bibinfo{year}{2003}\natexlab{}.
\newblock \showarticletitle{To Inline or Not to Inline? Enhanced Inlining
  Decisions}. In \bibinfo{booktitle}{\emph{Languages and Compilers for Parallel
  Computing, 16th International Workshop, {LCPC} 2003, College Station, TX,
  USA, October 2-4, 2003, Revised Papers}} \emph{(\bibinfo{series}{Lecture
  Notes in Computer Science}, Vol.~\bibinfo{volume}{2958})}.
  \bibinfo{publisher}{Springer}, \bibinfo{pages}{405--419}.
\newblock
\urldef\tempurl%
\url{https://doi.org/10.1007/978-3-540-24644-2\_26}
\showDOI{\tempurl}


\bibitem[Zuo et~al\mbox{.}(2018)]%
        {innereye}
\bibfield{author}{\bibinfo{person}{Fei Zuo}, \bibinfo{person}{Xiaopeng Li},
  \bibinfo{person}{Patrick Young}, \bibinfo{person}{Lannan Luo},
  \bibinfo{person}{Qiang Zeng}, {and} \bibinfo{person}{Zhexin Zhang}.}
  \bibinfo{year}{2018}\natexlab{}.
\newblock \showarticletitle{Neural machine translation inspired binary code
  similarity comparison beyond function pairs}.
\newblock \bibinfo{journal}{\emph{arXiv preprint arXiv:1808.04706}}
  (\bibinfo{year}{2018}).
\newblock


\end{thebibliography}



\end{document}